\documentclass[12pt]{article}
\usepackage[margin=25truemm]{geometry}
\usepackage{amsmath,amssymb}
\usepackage{physics}
\usepackage{mathtools}
\usepackage{bm}
\usepackage{bbm}
\usepackage{booktabs}
\usepackage{xcolor}
\usepackage{cite}

\newcommand{\dottedmidrule}{%
  \noalign{%
    \vskip\aboverulesep
    \hbox to \textwidth{%
      \leaders\hbox{\rule{0.4pt}{0.4pt}\hskip2pt}\hfill
    }%
    \vskip\belowrulesep
  }%
}

\numberwithin{equation}{section}

\begin{document}

\begin{titlepage}
\renewcommand{\thefootnote}{\fnsymbol{footnote}}
\begin{normalsize}
\begin{flushright}
\begin{tabular}{l}
November 2026
\end{tabular}
\end{flushright}
\end{normalsize}

~~\\

\vspace*{0cm}
    \begin{Large}
       \begin{center}
         {Matrix-vector models for the 3D Ising CFT }
       \end{center}
    \end{Large}
\vspace{1cm}

\begin{center}
           Kaichi M{\sc izohata}$^{1)}$\footnote
            {
e-mail address:
mizohata.kaichi.20@shizuoka.ac.jp},
           Tatsuya S{\sc eko}$^{1)}$\footnote
            {
e-mail address:
seko.tatsuya.18@shizuoka.ac.jp}
           {and}
           Asato T{\sc suchiya}$^{1),2)}$\footnote
           {
e-mail address: tsuchiya.asato@shizuoka.ac.jp}
\\
      \vspace{1cm}

$^{1)}$
 {\it Graduate School of Science and Technology, Shizuoka University}\\
               {\it 836 Ohya, Suruga-ku, Shizuoka 422-8529, Japan}\\
        \vspace{0.3cm}

 $^{2)}$
{\it Department of Physics, Shizuoka University}\\
                {\it 836 Ohya, Suruga-ku, Shizuoka 422-8529, Japan}\\

\end{center}

\vspace{1cm}

\begin{abstract}
\noindent
The construction of three-dimensional conformal field theories (3D CFTs) using
interacting fermions in the lowest Landau level on a sphere 
has recently proven remarkably efficient, while
it differs from 
that of 
conventional field theories on the fuzzy
sphere. 
Focusing on the 3D Ising CFT,  
we consider a class of models that are defined by
deformed Hamiltonians specified by a function which controls how the
interaction is damped at high angular momenta, 
and this class includes the
original model.
We show that each model in the class 
can be reformulated in a basis-independent manner 
as a matrix--vector quantum-mechanical
model, 
and, upon
integrating out the fermionic vectors, as a matrix model.
We then focus on a Gaussian choice, for which the bosonic matrices
acquire a heat-kernel propagator whose width is controlled by a parameter $\xi$. 
Through exact diagonalization with the number of fermions up to $16$, we find that
the low-energy spectra of the deformed Hamiltonians at criticality 
agree with those in the 3D Ising CFT for
a range of $\xi$. 
We thus obtain matrix-model realizations of the 3D Ising CFT that, in contrast
to conventional fuzzy-sphere field theories, show no sign of UV/IR mixing 
in the low-lying spectrum. 
\end{abstract}

\end{titlepage}
\vfil\eject

\setcounter{footnote}{0}

\tableofcontents

\section{Introduction} \label{intro}
Recent years have seen considerable progress in constructing
three-dimensional conformal field theories (3D CFTs) from systems of
interacting fermions confined to the lowest Landau level (LLL) 
on a sphere.
This approach is commonly called the fuzzy-sphere regularization of 3D CFTs.
The degeneracy of the LLL is $N=2s+1$, where $4 \pi s$ denotes the magnetic charge
of a monopole at the center of $S^2$.\footnote{In this paper, we only consider
positive $s$.} Through radial quantization, a 3D CFT on $\mathbb{R}^3$ is
formulated on $\mathbb{R}\times S^2$, 
so that a model defined by a 
Hamiltonian regularized on the sphere can provide the 3D CFT in the thermodynamic limit corresponding
to $N\to\infty$. An important advantage of this regularization is that it
preserves the $SO(3)$ rotational symmetry, a subgroup of the
three-dimensional conformal group $SO(4,1)$, which is difficult to maintain
in conventional lattice regularizations.
 
In the pioneering work \cite{Zhu:2022gjc}, a system of $N$ electrons at half
filling in the LLL was constructed to mimic the quantum Ising model on the
sphere. By tuning the parameters to the Ising transition, the authors showed
that the low-energy spectrum agrees with the conformal dimensions of
operators of the 3D Ising CFT determined by the conformal bootstrap,
signaling the emergence of conformal symmetry. Remarkably, the agreement is
already accurate for systems as small as $N=16$. Subsequent developments include 
studies of correlation functions and operator product expansion coefficients of the 3D Ising CFT \cite{Hu:2023xak, Han:2023yyb, Hofmann:2023llr, Lauchli:2025fii, Poland:2025ide, Fardelli:2026zas, Ghosh:2026xnp, Stergiou:2026ncu}, 
extensions to fractional electron fillings \cite{Voinea:2024ryq, Zhou:2026ugg}, 
studies of other 3D CFTs on $\mathbb{R} \times S^2$ \cite{Zhou:2023qfi, Chen:2023xjc, Han:2023lky, Chen:2024jxe, Zhou:2024zud, Yang:2025wqn, Fan:2025bhc, ArguelloCruz:2025zuq, EliasMiro:2025msj, He:2025ong, Taylor:2025odf, Yang:2025stu, Zhou:2025rmv, Voinea:2025iun, Zhou:2025kng, Dey:2025zgn, Guo:2025odn, Tang:2025wtj, Huffman:2026qqq, Eck:2026dzq, Sarma:2026lte, Dey:2026cso, Feng:2026iii, Stergiou:2026rir, DaLiao:2026gva, Camasca:2026awd, Yang:2026qtk, Liu:2026dlz, Baran:2026opo, Zhou:2026ugg, Qian:2026lrp, Tang:2026kzf, Dedushenko:2026zhy}, 
and extensions to other geometries \cite{Dedushenko:2024nwi, Zhou:2024dbt, Feng:2026iii, Dong:2025jjv, Meng:2026vcd, Qian:2026fbx, Meng:2026lfp, DaLiao:2026gva, Taylor:2026wan, Han:2025lrj}  among other developments \cite{Hu:2023ghk, Zhou:2023fqu, Hu:2024pen, Cuomo:2024psk, Fardelli:2024qla, Fan:2024vcz, Wiese:2025zfb, Hao:2026lnt, Zeng:2026cnh, Fardelli:2026ent}. (See \cite{Zhou:2025liv} for numerical implementation and \cite{He:2026ong} for review of the fuzzy sphere regularization.) 
 
This regularization differs from field theories on the
fuzzy sphere \cite{Madore:1991bw}, a representative noncommutative space, which are generally
described by matrix models. For example, the matrix quantum mechanics
\begin{align}
S=\int d\tau\,\frac{1}{N}\,\mathrm{Tr}\left(\frac12(\partial_\tau\Phi)^2
-\frac{1}{2 R^2}[L_i,\Phi]^2+\frac12m^2\Phi^2+\frac14g\Phi^4\right),
\label{fuzzy MQM}
\end{align}
where $\Phi(\tau)$ is an $N\times N$ Hermitian matrix and $L_i$ $(i=1,2,3)$
are the generators of the $N$-dimensional representation of $SU(2)$, defines
a scalar field theory on $\mathbb{R}\times S^2_{\rm fuzzy}$,
\begin{align}
S=\int d\tau\int_{S^2}\frac{d\Omega}{4\pi}\left(\frac12(\partial_\tau\phi)^2
-\frac{1}{2 R^2} (\mathcal{L}_i\phi)^2+\frac12m^2\phi^2+\frac14g\,\phi\star\phi\star\phi\star\phi\right),
\label{fuzzy field theory}
\end{align}
where $\mathcal{L}_i$ are the orbital angular momentum operators, $\star$ denotes the star product and $N$ acts as a UV cutoff. In the
large-$N$ limit, the classical action approaches that of the ordinary
$\phi^4$ theory on $\mathbb{R}\times S^2$, where $R$ corresponds to the radius of $S^2$. Quantum mechanically, however, the
two theories generally differ because of UV/IR mixing\footnote{It was shown
in \cite{Chu:2001xi} that the one-loop effective action in a scalar field theory 
on the fuzzy sphere differs from
that in the corresponding scalar field theory on an ordinary sphere by finite (non-divergent) amount even in the commutative
limit.
While this behavior is different from
IR divergence due to UV/IR mixing
originally found
in \cite{Minwalla:1999px}
and called noncommutative anomaly
in \cite{Chu:2001xi}, 
we widely refer to deviations in
quantum field theories 
on noncommutative spaces
from the corresponding 
ordinary quantum field theories including
the above phenomenon 
as UV/IR mixing 
in this paper.} 
arising from
nonplanar diagrams \cite{Minwalla:1999px,Chu:2001xi}\footnote{UV/IR mixing at non-perturbative level 
has been
studied by numerical simulations in \cite{Ambjorn:2002nj,Bietenholz:2004xs,Mejia-Diaz:2014lza,Bietenholz:2002ch,Bietenholz:2006cz,Martin:2004un,Panero:2006bx,GarciaFlores:2009hf,Das:2007gm,Hatakeyama:2018qjr,Hatakeyama:2017fao}.}.
 
By contrast, the LLL-based regularization yields ordinary field theories,
namely 3D CFTs, at criticality in the thermodynamic limit. This raises two related
questions. First, can such regularized theories be formulated as matrix
models, and if so, what distinguishes them from conventional fuzzy-sphere
field theories such as \eqref{fuzzy MQM}? Second, how much freedom is there
in the construction while still reaching the 3D CFT?
 
In this paper, to address these questions, we focus on the 3D Ising CFT and its LLL-based
regularization constructed in Ref.~\cite{Zhu:2022gjc}.
The Hamiltonian considered in Ref.~\cite{Zhu:2022gjc} involves 
a function
\begin{align}
\mathcal{F}_{\rm orig}\big(k(k+1),s\big)
=(2s+1)\begin{pmatrix} s & k & s\\ -s & 0 & s\end{pmatrix}^2
=\frac{(2s+1)!\,(2s)!}{(2s+k+1)!\,(2s-k)!}  \ ,
\label{F orig}
\end{align}
where $k$ stands for the angular momentum possessed by (spin) density operators, and
$( \;\;)$ is Wigner's $3j$ symbol.
It plays the role of a form factor that damps the interaction at angular momenta 
$k\gtrsim\sqrt{s}$. Since $k\mapsto k(k+1)$ is one-to-one for $k\ge0$, any
function of $k$, including \eqref{F orig}, can be regarded as a function of
$k(k+1)$, which allows it to be promoted to a function of the quadratic
Casimir operator, as we will see. We thus consider a class of models that are defined by
Hamiltonians obtained by replacing $\mathcal{F}_{\rm orig}$ with a general function $\mathcal{F}$. This class includes
the original model, and these deformations preserve all symmetries possessed by the original model.
We show that each model in the class 
can be reformulated 
as a matrix--vector quantum-mechanical
model, 
and, upon
integrating out the fermionic vectors, as a matrix model.
The matrix--vector models and the matrix models obtained here are
written in a basis-independent form. The actions consist only of traces,
commutators with $L_i$, and functions of $(L^{\rm ad}_i)^2$, where $L^{\rm ad}_iX=[L_i,X]$.
The geometry of the sphere enters only through the
background $L_i$. This makes it natural to regard $L_i$ as a background
of a more fundamental matrix model, 
and
suggests that other backgrounds may be considered in the same framework.
Although we restrict ourselves to the 3D Ising CFT in this
paper, the reformulation relies only on the Hamiltonian consisting of
LLL-projected density--density interactions, and its extension to LLL
models for other 3D CFTs is straightforward, as we discuss in section 5.

Among the possible choices of $\mathcal{F}$, as an example, we  
focus on the Gaussian form,
which is simpler than $\mathcal{F}_{\rm orig}$,
\begin{align}
\mathcal{F}_\xi\big(k(k+1),s\big)=\exp\left(-\frac{\xi\,k(k+1)}{2s}\right),
\label{F Gaussian}
\end{align}
for which the bosonic matrices acquire a heat-kernel type propagator
proportional to
$e^{-\xi k(k+1)/2s}$ in the basis of fuzzy spherical harmonics. For
$k\ll s^{3/4}$, one finds
$\mathcal{F}_{\rm orig}=\mathcal{F}_{\xi=1}\,[1+O(k^2/s^2,\,k^4/s^3)]$, so
that at $\xi=1$ the Gaussian model coincides with the original one up to
corrections that are $O(1/s)$ in the relevant range $k=O(\sqrt{s})$. For
$\xi\neq1$, in contrast, $\sup_k|\mathcal{F}_\xi-\mathcal{F}_{\rm orig}|$
remains finite as $s\to\infty$, and the model differs from the original one
even in the thermodynamic limit. 
By exact diagonalization for $N\le16$, we
find that for 
a range of $\xi$ the low-energy spectra at criticality agree
with the 3D Ising CFT to an accuracy comparable to that of the original
model, and that the deviations decrease as $N$ increases. For the Gaussian
deformation of a simplified Hamiltonian \cite{He:2025ong}, whose
matrix-model formulation contains fewer matrices, the agreement holds for
a narrower range of $\xi$,
which shows that the agreement is not automatic.
 
These results provide matrix-model realizations of the 3D Ising CFT. Because
they reproduce the CFT spectrum, these matrix models do not suffer from
UV/IR mixing, at least at criticality and for the low-lying spectrum
examined here. The matrix-model formulation also opens the possibility of
studying these models at larger $N$ by Monte Carlo methods.
The structural difference from \eqref{fuzzy MQM} is
instructive. In \eqref{fuzzy MQM}, the field itself is a matrix whose modes
up to $k=2s$ are all dynamical. In our models, the bosonic matrices are
auxiliary fields whose propagator suppresses modes with
$k\gtrsim\sqrt{s/\xi}$, and the low-energy degrees of freedom are
collective excitations of fermions in the fundamental representation. This
suggests a mechanism for obtaining ordinary quantum field theories from
matrix models that differs both from large-$N$ reduction \cite{EK},
which is restricted to large-$N$ theories, and from field theories on
noncommutative spaces \cite{Gonzalez-Arroyo:1982hyq,Aoki:1999vr}, which deviate from 
ordinary quantum field theories due to UV/IR mixing. It also raises the
question of whether models of this type emerge as low-energy effective
theories of more fundamental matrix models, such as the BFSS
\cite{BFSS}, BMN \cite{BMN}, and type IIB \cite{IKKT} matrix
models, after integrating out ultraviolet degrees of freedom around a
background characterized by $L_i$. In this respect, it is
important to find which
choices of $\mathcal{F}$ yield
the 3D Ising model,
and it is suggestive that
the heat-kernel propagator in our models is reminiscent of the exact
renormalization group \cite{Polchinski:1983gv}, with $2s/\xi$ playing the role of an effective scale. 
The fermionic vectors can also be
embedded as off-diagonal blocks of larger matrices.
 
The paper is organized as follows. In section 2, we review conformal field
theory, the Landau problem on the sphere, and the LLL-based fuzzy-sphere
regularization of the 3D Ising CFT. In section 3, we introduce the class of
Hamiltonians specified by $\mathcal{F}$ and reformulate them as vector models,
matrix--vector models, and matrix models, for both the original and the
simplified Hamiltonians. In section 4, we focus on the Gaussian choice
\eqref{F Gaussian} and examine the spectra of the resulting models by exact
diagonalization, showing that they are consistent with the 3D Ising CFT for
a range of $\xi$. Section 5 is devoted to conclusions and discussion.
The appendices provide details of the parameter tuning and of the numerical
results.

\section{Review} \label{review}

In this section, we review the elements of conformal field theory used in
this paper, the Landau problem on the sphere, the LLL-based fuzzy-sphere
regularization of the 3D Ising CFT \cite{Zhu:2022gjc}, and its simplified
version \cite{He:2025ong}.

\subsection{CFT} \label{sec:review_CFT}

Let us consider a CFT in $d$-dimensional Euclidean space. Its conformal group
$SO(d+1,1)$ is generated by translations $P_\mu$, rotations $M_{\mu\nu}$,
dilatations $D$, and special conformal transformations $K_\mu$, where
$\mu,\nu=1,\ldots,d$. Since $[M_{\mu\nu},D]=0$, operators can be classified
by the eigenvalue $\Delta$ of $D$ acting on operators at $x^\mu=0$, called
the conformal dimension, and by irreducible representations of the rotation
group $SO(d)$. Primary operators at $x^\mu=0$ are annihilated by
$K_\mu$.
Descendants are obtained by acting repeatedly with $P_\mu$ on
primary operators, and each primary operator and its descendants form a
conformal multiplet.

In three dimensions, the integer-spin irreducible representations of
$SO(3)$ are described by rank-$\ell$ symmetric traceless tensors, and each
operator is characterized by $(\Delta,\ell)$. If the CFT is invariant under
spacetime parity, as is the 3D Ising CFT, the rotation group is extended to
$O(3)$, and each operator also carries a parity $P=\pm1$. Operators may
also carry charges of internal symmetries, such as the $\mathbb{Z}_2$ charge
in the 3D Ising CFT. In the 3D Ising CFT, we refer to the set of operators with given
$(\mathbb{Z}_2,P,\ell)$ as a sector.  In the expressions
below, free indices are understood to be symmetrized and made traceless,
while repeated indices are contracted. The descendants of a scalar primary
operator $O$ ($\ell=0$) take the form
\begin{equation}  \partial_{\nu_1}\cdots\partial_{\nu_j}\Box^n O\ , \quad (n,j\ge0)\ ,
  \label{O_descendant_1}
\end{equation}
where $\Box=\partial^2$, and have quantum numbers $(\Delta+2n+j,j)$. The
descendants of a primary operator $O_{\mu_1\cdots\mu_\ell}$ with spin
$\ell\ge1$ are of two types. The first type takes the form
\begin{equation}
\partial_{\nu_1}\cdots\partial_{\nu_j}\partial_{\mu_1}\cdots\partial_{\mu_i}
  \Box^n O_{\mu_1\cdots\mu_\ell}\ , \quad (\ell\ge i\ge0,\ n,j\ge0)\ ,
  \label{O_descendant_2}
\end{equation}
with quantum numbers $(\Delta+2n+j+i,\ \ell+j-i)$. The second type involves
the Levi-Civita tensor $\epsilon_{\mu\nu\rho}$,
\begin{equation}
\epsilon_{\tau\rho\mu_\ell}\,\partial_\rho\,\partial_{\nu_1}\cdots\partial_{\nu_j}  \partial_{\mu_1}\cdots\partial_{\mu_i}\Box^n O_{\mu_1\cdots\mu_\ell}\ , \quad
  (\ell-1\ge i\ge0,\ n,j\ge0)\ ,
  \label{O_descendant_3}
\end{equation}
with quantum numbers $(\Delta+2n+j+i+1,\ \ell+j-i)$. The descendants
\eqref{O_descendant_1} and \eqref{O_descendant_2} have the same parity as
the primary operator, whereas the factor $\epsilon_{\mu\nu\rho}$ in
\eqref{O_descendant_3} flips it. For conserved currents, such as the
energy-momentum tensor, only $i=0$ appears in \eqref{O_descendant_2} and
\eqref{O_descendant_3}. 
A spin-$\ell$  operator $O_{\mu_1\cdots\mu_\ell}$
(primary or descendant)
has $2\ell+1$ independent components, which can be labeled by the
eigenvalue $L_z=-\ell,\ldots,\ell$ of the third component of the angular
momentum, and exactly one of them has $L_z=0$. In this paper, we count
$O_{\mu_1\cdots\mu_\ell}$ as one operator rather than as $2\ell+1$
operators. The number of operators with a given $\Delta$ in a given sector
is thus equal to the number of their $L_z=0$ components, and it suffices to
consider the $L_z=0$ components.

We next recall the state-operator correspondence. Under a Weyl
transformation, $\mathbb{R}^d$ is mapped to the cylinder
$\mathbb{R}\times S^{d-1}$ with radius $R$. In a lattice or fuzzy-sphere
model, the emergent conformal symmetry involves a model-dependent velocity
$v$, which plays the role of the speed of light. The radial coordinate $r$
of $\mathbb{R}^d$ is then related to the Euclidean time $\tau$ of the model
by
\begin{equation}
  \tau=\frac{R}{v}\log\frac{r}{R}\ .
  \label{radial_time}
\end{equation}
The dilatation $r\to e^\lambda r$ thus corresponds to the time translation
$\tau\to\tau+R\lambda/v$, and $D$ corresponds to $(R/v)H$, where $H$ is the
Hamiltonian on $\mathbb{R}\times S^{d-1}$. States annihilated by $K_\mu$ are called primary
states, and states obtained by acting with $P_\mu$ on them are called
descendant states. There is a one-to-one correspondence between operators on
$\mathbb{R}^d$ and states on $\mathbb{R}\times S^{d-1}$, under which the
excitation energy $\Delta E_n=E_n-E_0$ of an eigenstate $\ket{\psi_n}$ is
proportional to the conformal dimension $\Delta_n$ of the corresponding
operator,
\begin{equation}
  \Delta E_n=\frac{v}{R}\,\Delta_n\ ,
  \label{state-operator}
\end{equation}
which follows from $D=(R/v)H$. The rotation group of $S^{d-1}$ is identified with that of $\mathbb{R}^d$, so
that the spin and parity of $\ket{\psi_n}$ agree with those of the
corresponding operator, and quantum numbers of internal symmetries, such as
the $\mathbb{Z}_2$ of the Ising CFT, are likewise preserved.

Since $v$ is not known a priori in a lattice or fuzzy-sphere model, we fix
the normalization in \eqref{state-operator} by the energy-momentum tensor
$T_{\mu\nu}$, whose conformal dimension is exactly $3$. In the 3D Ising CFT,
$T_{\mu\nu}$ is the lowest operator with $(\mathbb{Z}_2,P,\ell)=(+,+,2)$. We
rescale all excitation energies so that the lowest state in this sector has
$\Delta=3$, and refer to the rescaled excitation energies as (numerical)
conformal dimensions.

\subsection{Landau problem on the sphere} \label{sec:review_Landau}
Let us consider an electron of mass $M_e$ moving on a sphere of radius $R$ with a magnetic monopole of charge 
$4\pi s$ $(2s\in\mathbb{Z})$ at its 
center. The
monopole produces a uniform magnetic field of strength $B$ on the sphere,
and we define the magnetic length $\ell_B$ by
\begin{equation}
  B=\frac{s}{R^2}\ , \qquad \ell_B=\frac{1}{\sqrt{B}}=\frac{R}{\sqrt{s}}\ ,
  \label{B_lB}
\end{equation}
in units with $\hbar=e=c=1$. Let $x_i$ $(i=1,2,3)$ be the Cartesian
coordinates of $\mathbb{R}^3$ in which the sphere is embedded. The
Hamiltonian is
\begin{equation}
  H_0=\frac{1}{2M_eR^2}\Lambda_i^2\ ,
  \label{H_0_withLambda}
\end{equation}
where
$\Lambda_i=\varepsilon_{ijk}x_j(-i\partial_k+A_k)$ are the covariant orbital
angular momenta and $A_i$ is the monopole gauge field. The angular momenta
that commute with $H_0$ are
\begin{equation}
  L_i=\Lambda_i+s\,\frac{x_i}{R}\ ,
  \label{L}
\end{equation}
where the second term is the contribution of the electromagnetic field. They
satisfy $[L_i,L_j]=i\varepsilon_{ijk}L_k$ and $L_i^2=\Lambda_i^2+s^2$, so that
\begin{equation}
  H_0=\frac{1}{2M_eR^2}\left(L_i^2-s^2\right)\ .
  \label{H_with_L}
\end{equation}
The eigenstates of \eqref{H_with_L} form Landau levels. The $n$-th Landau
level has energy
\begin{equation}
  E_n=\frac{1}{2M_eR^2}\big(n(n+1)+(2n+1)s\big)\ , \quad (n=0,1,2,\ldots)\ ,
\end{equation}
and degeneracy $2s+2n+1$, the dimension of the spin-$(s+n)$
representation.

The LLL ($n=0$) has degeneracy $N=2s+1$. 
Its wave functions are the monopole harmonics
\begin{equation}
  Y^{(s)}_{s,m}(\Omega)
=N_m\,e^{im\varphi}\cos^{s+m}\!\left(\frac{\theta}{2}\right)
  \sin^{s-m}\!\left(\frac{\theta}{2}\right)\ , \quad (m=-s,-s+1,\ldots,s)\ ,
  \label{monopole_harmonics}
\end{equation}
where $\Omega=(\theta,\varphi)$ are the angular coordinates, 
and
$N_m=\sqrt{(2s+1)!/\big(4\pi(s+m)!(s-m)!\big)}$. Each of $Y^{(s)}_{s,m}(\Omega)$ is localized near
a particular latitude on the sphere as seen 
in Fig.~\ref{fig:phase_diagram}.  
The monopole harmonics form
the spin-$s$ irreducible representation of $SO(3)$:
\begin{equation}
L_i^2\,Y^{(s)}_{s,m}=s(s+1)\,Y^{(s)}_{s,m}\ , \qquad
  L_3\,Y^{(s)}_{s,m}=m\,Y^{(s)}_{s,m}\ .
\end{equation}
For a review of the relation among monopoles, the LLL and fuzzy spheres, see \cite{Hasebe:2010vp}. In particular, projections of the coordinates $x^i$ of the sphere onto the monopole harmonics 
\begin{align}
(X^i)_{m_1,m_2} =\int d\Omega  \; Y_{s,m_1}^{(s)*}  \; x^i \; Y_{s,m_2}^{(s)}
\end{align}
satisfy
\begin{align}
[X^i,X^j]=\frac{R}{s+1}i\epsilon^{ijk}X^k,  \;\;\; (X^i)^2=\frac{s R^2}{s+1} \mathbbm{1}  \ .
\end{align}
The $(2s+1)\times (2s+1)$ matrices $X^i$ are viewed as coordinates of the fuzzy sphere \cite{Ishiki:2018aja}\footnote{This is an example of the Brezin-Toeplitz quantization\cite{Klimek1992, Bordemann:1993zv, Ma2008}.}. This shows a relation between the fuzzy sphere regularization reviewed in the next
subsection and noncommutative geometry\cite{Zhu:2022gjc,He:2026ong}. 

The product $Y^{(s)*}_{s,m_1}\,Y^{(s)}_{s,m_4}$ has zero monopole
charge and can be expanded in the ordinary spherical harmonics $Y_{k,m}$
with $k=0,1,\ldots,2s$. The expansion coefficients are
\begin{align}
  \int_{S^2}d\Omega\ Y^{(s)*}_{s,m_1}(\Omega) \,Y^{(s)}_{s,m_4}(\Omega)\,
  Y_{k,m}(\Omega)
  &=(-1)^{s-m_1 + k}(2s+1)\sqrt{\frac{2k+1}{4\pi}}\nonumber\\
  &\quad\times
  \begin{pmatrix} s & k & s\\ -m_1 & m & m_4\end{pmatrix}
  \begin{pmatrix} s & k & s\\ -s & 0 & s\end{pmatrix}\ ,
  \label{overlap}
\end{align}
where $d\Omega=\sin\theta\,d\theta\,d\varphi$.
The second $3j$ symbol is independent of $m_1$ and $m_4$, and
it is the origin of the function $\mathcal{F}_{\rm orig}$ in \eqref{F orig}.

\subsection{Fuzzy-sphere regularization of the 3D Ising CFT} \label{sec:review_Ising}

We now review the fuzzy-sphere regularization of the 3D Ising CFT
constructed in Ref.~\cite{Zhu:2022gjc}. We consider electrons on the sphere
carrying a pseudospin $\alpha=\uparrow,\downarrow$, with the Hamiltonian
\begin{equation}
\hat{H}=\frac12\int_{S^2}d\Omega_a\,d\Omega_b\ U(\Omega_{ab})
\left(\hat{n}^0(\Omega_a)\hat{n}^0(\Omega_b)-\hat{n}^z(\Omega_a)\hat{n}^z(\Omega_b)\right)
  -h\int_{S^2}d\Omega\ \hat{n}^x(\Omega)\ ,
  \label{H_original_psi}
\end{equation}
where
\begin{equation}
  \hat{n}^a(\Omega)=
\left(\hat{\psi}^\dagger_\uparrow(\Omega)\ ,\ \hat{\psi}^\dagger_\downarrow(\Omega)\right)
  \sigma^a
\begin{pmatrix}\hat{\psi}_\uparrow(\Omega)\\ \hat{\psi}_\downarrow(\Omega)\end{pmatrix}
  \quad (a=0,x,y,z)
  \label{na}
\end{equation}
with $\sigma^{x,y,z}$ the Pauli matrices and $\sigma^0=I_{2\times2}$. Here
$\hat{n}^0$ is the number density and $\hat{n}^{x,y,z}$ are the pseudospin
densities. The fermion operators are projected onto the LLL,
\begin{equation}
  \hat{\psi}_\alpha(\Omega)=\sum_{m=-s}^{s}Y^{(s)}_{s,m}(\Omega)\,\hat{c}_{m,\alpha}\ ,
  \label{psi_LLL}
\end{equation}
where $\hat{c}_{m,\alpha}$ annihilates an electron with pseudospin $\alpha$
in the orbital $m$ and obeys
$\{\hat{c}_{m,\alpha},\hat{c}_{n,\beta}^{\dagger}\}=\delta_{mn}\delta_{\alpha\beta}$. 
The interaction is
\begin{equation}
  U(\Omega_{ab})=4\pi\left(\frac{g_0}{R^2}\,\delta(\Omega_{ab})
  +\frac{g_1}{R^4}\,\nabla^2\delta(\Omega_{ab})\right)\ ,
  \label{U}
\end{equation}
where $\delta(\Omega_{ab})$ is the delta function on the unit sphere,
normalized as $\int d\Omega_a\,\delta(\Omega_{ab})=1$, and $\nabla^2$ is the
Laplacian on the unit sphere. 
This system can be viewed as a
mimic of the quantum
Ising model on the sphere.
Expanding $U$ in Legendre polynomials as
$U(\Omega_{ab})=\sum_k U_k\,P_k(\cos\theta_{ab})$, where $\theta_{ab}$ is the
angle between $\Omega_a$ and $\Omega_b$, gives
\begin{equation}
  U_k=\frac{g_0}{R^2}(2k+1)+\frac{g_1}{R^4}(2k+1)\big(-k(k+1)\big)\ .
  \label{U_k}
\end{equation}
With $\ell_B$ fixed, $R\propto\sqrt{s}$ by \eqref{B_lB}, and the
thermodynamic limit corresponds to $s\to\infty$. In numerical calculations,
we set $R^2=N=2s+1$, so that $g_0$ and $g_1$ remain of order one. Of the
three parameters $h$, $g_0$, and $g_1$, one can be fixed by the choice of
energy unit, leaving two dimensionless parameters.

Substituting \eqref{psi_LLL}
into \eqref{H_original_psi}, 
we obtain
\begin{align}
  \hat{H}&=\frac12\sum_{m_{1,2,3,4}=-s}^{s}V_{m_1,m_2,m_3,m_4}  \left(\big(\hat{\bm{c}}^\dagger_{m_1}\hat{\bm{c}}_{m_4}\big)
\big(\hat{\bm{c}}^\dagger_{m_2}\hat{\bm{c}}_{m_3}\big)
  -\big(\hat{\bm{c}}^\dagger_{m_1}\sigma^z\hat{\bm{c}}_{m_4}\big)  \big(\hat{\bm{c}}^\dagger_{m_2}\sigma^z\hat{\bm{c}}_{m_3}\big)\right)\nonumber\\
  &\quad-h\sum_{m=-s}^{s}\hat{\bm{c}}^\dagger_m\sigma^x\hat{\bm{c}}_m\ ,
  \label{H_original_c}
\end{align}
where $\hat{\bm{c}}_m=(\hat{c}_{m,\uparrow},\hat{c}_{m,\downarrow})^{\rm T}$.
Using the addition theorem
\begin{equation}
  P_k(\cos\theta_{ab})=\frac{4\pi}{2k+1}\sum_{m=-k}^{k}Y_{k,m}(\Omega_a)\ Y_{k,m}^*(\Omega_b)
  \nonumber
\end{equation}
and \eqref{overlap}, we obtain
\begin{align}
  V_{m_1,m_2,m_3,m_4}
  &=\int d\Omega_a\,d\Omega_b\ U(\Omega_{ab})\,
  Y^{(s)*}_{s,m_1}(\Omega_a)\,Y^{(s)}_{s,m_4}(\Omega_a)\,
  Y^{(s)*}_{s,m_2}(\Omega_b)\,Y^{(s)}_{s,m_3}(\Omega_b)\nonumber\\
  &=\sum_{k=0}^{2s}U_k\,(-1)^{2s+m_2+m_4}
  (2s+1)
  \begin{pmatrix} s & k & s\\ -m_1 & m_1-m_4 & m_4\end{pmatrix}
  \begin{pmatrix} s & k & s\\ -m_2 & m_2-m_3 & m_3\end{pmatrix}\nonumber\\
  &\quad\quad\times
    \mathcal{F}_{\mathrm{orig}}(k(k+1),s) \ \delta_{m_1+m_2,m_3+m_4}\ ,
  \label{V1234}
\end{align}
where $\mathcal{F}_{\mathrm{orig}}$ is given in \eqref{F orig}.

The coefficients \eqref{V1234} can also be expressed in terms of the Haldane
pseudopotentials $V_l$ \cite{Haldane:1983xm}, which give the interaction
energy of a pair of electrons with relative angular momentum $l$, that is,
with total angular momentum $2s-l$:
\begin{align}
  V_{m_1,m_2,m_3,m_4}&=\sum_{l=0}^{2s}V_{2s-l}\,(2l+1)
  \begin{pmatrix} s & s & l\\ m_1 & m_2 & -m_1-m_2\end{pmatrix}
  \begin{pmatrix} s & s & l\\ m_4 & m_3 & -m_3-m_4\end{pmatrix}\nonumber\\
  &\quad\times\delta_{m_1+m_2,m_3+m_4}
  \label{V1234_with_Haldane_potential}
\end{align}
with
\begin{equation}
  V_{2s-l}=\sum_{k=0}^{2s}U_k\,(-1)^{2s+l}
  (2s+1)
  \begin{Bmatrix} l & s & s\\ k & s & s\end{Bmatrix}
  \mathcal{F}_{\mathrm{orig}}(k(k+1),s) \ ,
  \label{Haldane_potential}
\end{equation}
where $\{\cdots\}$ is Wigner's $6j$ symbol. For $U_k$ in \eqref{U_k},
$V_l=0$ for $2\le l\le 2s$, so that $h$, $V_0$, and $V_1$ can be used as the
parameters instead of $h$, $g_0$, and $g_1$.

We consider $N=2s+1$ electrons. Since there are $2(2s+1)$ single-particle
states including pseudospin, this corresponds to half filling. The
Hamiltonian possesses the following three symmetries.
\begin{enumerate}
  \item The Ising $\mathbb{Z}_2$ symmetry:
  \begin{equation}
    \begin{pmatrix}\hat{c}_{m,\uparrow}\\ \hat{c}_{m,\downarrow}\end{pmatrix}
    \to\sigma^x
    \begin{pmatrix}\hat{c}_{m,\uparrow}\\ \hat{c}_{m,\downarrow}\end{pmatrix}\ .
    \label{Z2 sym}
  \end{equation}
  \item The particle--hole symmetry:
  \begin{equation}
    \begin{pmatrix}\hat{c}_{m,\uparrow}\\ \hat{c}_{m,\downarrow}\end{pmatrix}
    \to i\sigma^y
    \begin{pmatrix}\hat{c}^\dagger_{m,\uparrow}\\ \hat{c}^\dagger_{m,\downarrow}\end{pmatrix}\ .
    \label{PH sym}
  \end{equation}
  \item The $SO(3)$ rotation symmetry:
  \begin{equation}
    \hat{c}_{m,\alpha}\to\sum_{m'=-s}^{s}D^{(s)}_{mm'}(g)\,\hat{c}_{m',\alpha}\ ,
    \quad g\in SO(3)\ ,
    \label{SO(3) sym}
  \end{equation}
  where $D^{(s)}(g)$ is the Wigner matrix of the spin-$s$ representation.
  Through \eqref{psi_LLL}, this is the rotation
  $\hat{\psi}_\alpha(\Omega)\to\hat{\psi}_\alpha(g^{-1}\Omega)$, up to a gauge
  transformation.
\end{enumerate}
These ultraviolet symmetries correspond to symmetries of the
3D Ising CFT in the infrared. The correspondence for the Ising $\mathbb{Z}_2$
and the rotation symmetry is evident. The particle--hole symmetry is
identified with spacetime parity. To see this, consider the $SO(3)$ vector
\begin{equation}
  n^x_{M}=\sum_{m_1=-s}^{s}(-1)^{s-m_1}
  \begin{pmatrix} s & s & 1\\ m_1 & M-m_1 & -M\end{pmatrix}
  \hat{\bm{c}}^\dagger_{m_1}\sigma^x\hat{\bm{c}}_{m_1-M}\ , \quad (M=0,\pm1)\ .
  \nonumber
\end{equation}
Under \eqref{PH sym}, it transforms as $n^x_M\to(-1)^M n^x_{-M}$:
\begin{equation}
  \begin{pmatrix} n^x_{1}\\ n^x_{0}\\ n^x_{-1}\end{pmatrix}
  \to
  \begin{pmatrix} 0 & 0 & -1\\ 0 & 1 & 0\\ -1 & 0 & 0\end{pmatrix}
  \begin{pmatrix} n^x_{1}\\ n^x_{0}\\ n^x_{-1}\end{pmatrix}\ .
\end{equation}
In Cartesian components, this is the reflection $n^x_2\to-n^x_2$, that is,
the spatial inversion combined with a rotation by $\pi$ about the 2-axis.
Thus the particle--hole symmetry acts as the improper $\mathbb{Z}_2$ of
$O(3)$. The energy eigenstates are therefore labeled by the energy $E$
and the sector $(\mathbb{Z}_2,P,\ell)$ defined in section \ref{sec:review_CFT}.

\begin{figure}[tbp]
  \centering
  \includegraphics[width=0.6\linewidth]{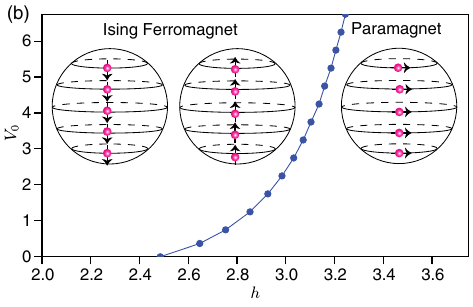}
  \caption{Phase diagram at $V_1=1$.
  Reproduced from Fig.~2 of W.~Zhu, C.~Han, E.~Huffman,
  J.~S.~Hofmann, and Y.-C.~He,
  ``Uncovering Conformal Symmetry in the 3D Ising Transition:
  State-Operator Correspondence from a Quantum Fuzzy Sphere
  Regularization,''
  Phys.\ Rev.\ X \textbf{13}, 021009 (2023),
  doi:10.1103/PhysRevX.13.021009
  (Ref.~\cite{Zhu:2022gjc}).
  Cropped; licensed under CC BY 4.0.}
  \label{fig:phase_diagram}
\end{figure}

For $h=0$ and $V_0,V_1>0$, the ground state is the Ising ferromagnet, which
breaks the Ising $\mathbb{Z}_2$ spontaneously and is doubly degenerate:
\begin{equation}
  \ket{\Psi_\uparrow}=\prod_{m=-s}^{s}\hat{c}^\dagger_{m,\uparrow}\ket{0}\ , \qquad
  \ket{\Psi_\downarrow}=\prod_{m=-s}^{s}\hat{c}^\dagger_{m,\downarrow}\ket{0}\ .
\end{equation}
For $h\gg V_0,V_1$, the ground state is the paramagnet, which preserves the
Ising $\mathbb{Z}_2$:
\begin{equation}
  \ket{\Psi_x}=\prod_{m=-s}^{s}\frac{1}{\sqrt2}
  \left(\hat{c}^\dagger_{m,\uparrow}+\hat{c}^\dagger_{m,\downarrow}\right)\ket{0}\ .
\end{equation}
An Ising transition therefore occurs as $h$ increases. The phase diagram in
units with $V_1=1$ is shown in Fig.~\ref{fig:phase_diagram}.
Ref.~\cite{Zhu:2022gjc} demonstrated a one-to-one correspondence, up
to finite-size corrections, between the low-energy eigenstates of
\eqref{H_original_c} at the transition and the operators of the 3D Ising
CFT. The model \eqref{H_original_c} thus provides a regularization of the
3D Ising CFT. We refer to it as the original model.

\subsection{Simplified model} \label{sec:review_simple}

The simplified model \cite{He:2025ong} is defined by
\begin{equation}
  \hat{H}_s=4\pi\int_{S^2}d\Omega
  \left(\frac12\frac{g_1}{R^4}\big(\nabla_i\hat{n}^z\big)^2
  +\frac12\frac{g_0}{R^2}\big(\hat{n}^0\big)^2\right)
  -h\int_{S^2}d\Omega\ \hat{n}^x\ ,
  \label{H_simple_3j_psi}
\end{equation}
which is obtained from \eqref{H_original_psi} by dropping
\begin{equation}
  4\pi\int_{S^2}d\Omega
  \left(-\frac12\frac{g_1}{R^4}\big(\nabla_i\hat{n}^0\big)^2
  -\frac12\frac{g_0}{R^2}\big(\hat{n}^z\big)^2\right)\ .
  \label{H_simple_gradient}
\end{equation}
In terms of the orbital operators, 
the Hamiltonian \eqref{H_simple_3j_psi}
reads
\begin{align}
  \hat{H}_s
  &=\frac12\sum_{m_{1,2,3,4}=-s}^{s}
  \left(V^0_{m_1,m_2,m_3,m_4}
\big(\hat{\bm{c}}^\dagger_{m_1}\hat{\bm{c}}_{m_4}\big) \big(\hat{\bm{c}}^\dagger_{m_2}\hat{\bm{c}}_{m_3}\big)
  -V^1_{m_1,m_2,m_3,m_4}
\big(\hat{\bm{c}}^\dagger_{m_1}\sigma^z\hat{\bm{c}}_{m_4}\big)
\big(\hat{\bm{c}}^\dagger_{m_2}\sigma^z\hat{\bm{c}}_{m_3}\big)\right)\nonumber\\
  &\quad-h\sum_{m=-s}^{s}\hat{\bm{c}}^\dagger_m\sigma^x\hat{\bm{c}}_m\ ,
  \label{H_simple_3j_c}
\end{align}
where $V^i_{m_1,m_2,m_3,m_4}$ $(i=0,1)$ are obtained from \eqref{V1234} by
replacing $U_k$ with
\begin{equation}
  U^0_k=\frac{g_0}{R^2}(2k+1)\ , \qquad
  U^1_k=\frac{g_1}{R^4}(2k+1)\big(-k(k+1)\big)\ ,
  \label{Uki}
\end{equation}
that is,
\begin{align}
  V^i_{m_1,m_2,m_3,m_4}
  &=\sum_{k=0}^{2s}U^i_k\,(-1)^{2s+m_2+m_4}(2s+1)
  \begin{pmatrix} s & k & s\\ -m_1 & m_1-m_4 & m_4\end{pmatrix}
  \begin{pmatrix} s & k & s\\ -m_2 & m_2-m_3 & m_3\end{pmatrix}\nonumber\\
  &\quad\times\mathcal{F}_{\rm orig}\big(k(k+1),s\big)\,\delta_{m_1+m_2,m_3+m_4}\ .
  \label{V1234i}
\end{align}
Thus the two channels have different interactions $U^0_k$ and $U^1_k$ but
share the same function $\mathcal{F}_{\rm orig}$. We consider the half-filling case. The simplified model
possesses all three symmetries \eqref{Z2 sym}, \eqref{PH sym}, and
\eqref{SO(3) sym}. At criticality, its low-energy
spectrum agrees with the 3D Ising CFT \cite{He:2025ong}, which we confirm
independently in section 4.

\section{Matrix--vector models for a class of Hamiltonians} \label{matrix-vector model}

In this section, we introduce the class of Hamiltonians specified by a
function $\mathcal{F}$. This class
includes the original and simplified models
reviewed in sections \ref{sec:review_Ising} and \ref{sec:review_simple}.
We reformulate every model in this class as a vector model, a matrix--vector
model, and a matrix model.

\subsection{A class of Hamiltonians} \label{sec:F_class}

Let $\mathcal{F}(k(k+1),s)$ be a function that is positive for
$k=0,1,\ldots,2s$. We define
\begin{align}
  \hat{H}_{\mathcal{F}}
  &=\frac12\sum_{m_{1,2,3,4}=-s}^{s}V^{\mathcal{F}}_{m_1,m_2,m_3,m_4}
  \left(\big(\hat{\bm{c}}^\dagger_{m_1}\hat{\bm{c}}_{m_4}\big)
  \big(\hat{\bm{c}}^\dagger_{m_2}\hat{\bm{c}}_{m_3}\big)
  -\big(\hat{\bm{c}}^\dagger_{m_1}\sigma^z\hat{\bm{c}}_{m_4}\big)
  \big(\hat{\bm{c}}^\dagger_{m_2}\sigma^z\hat{\bm{c}}_{m_3}\big)\right)\nonumber\\
  &\quad-h\sum_{m=-s}^{s}\hat{\bm{c}}^\dagger_m\sigma^x\hat{\bm{c}}_m\ ,
  \label{H_F_c}
\end{align}
where $V^{\mathcal{F}}_{m_1,m_2,m_3,m_4}$ is 
obtained by replacing 
$\mathcal{F}_{\mathrm{orig}}$
in \eqref{V1234} with
$\mathcal{F}$:
\begin{align} V_{m_1,m_2,m_3,m_4}^{\mathcal{F}}&=\sum_{k=0}^{2s}U_k\,(-1)^{2s+m_2+m_4}(2s+1)
\begin{pmatrix} s & k & s\\ -m_1 & m_1-m_4 & m_4\end{pmatrix}
\begin{pmatrix} s & k & s\\ -m_2 & m_2-m_3 & m_3\end{pmatrix}\nonumber\\
 &\quad\times\mathcal{F}\big(k(k+1),s\big)\,\delta_{m_1+m_2,m_3+m_4}  \ .
 \label{V general}
 \end{align}
For $\mathcal{F}=\mathcal{F}_{\rm orig}$ in
\eqref{F orig}, $\hat{H}_{\mathcal{F}}$ reduces to the original Hamiltonian
\eqref{H_original_c}. Similarly, we define $\hat{H}_{s,\mathcal{F}}$ by
\begin{align}
  \hat{H}_{s,\mathcal{F}}
  &=\frac12\sum_{m_{1,2,3,4}=-s}^{s}
\left(V^{\mathcal{F},0}_{m_1,m_2,m_3,m_4} \big(\hat{\bm{c}}^\dagger_{m_1}\hat{\bm{c}}_{m_4}\big)  \big(\hat{\bm{c}}^\dagger_{m_2}\hat{\bm{c}}_{m_3}\big)
  -V^{\mathcal{F},1}_{m_1,m_2,m_3,m_4}  \big(\hat{\bm{c}}^\dagger_{m_1}\sigma^z\hat{\bm{c}}_{m_4}\big) \big(\hat{\bm{c}}^\dagger_{m_2}\sigma^z\hat{\bm{c}}_{m_3}\big)\right)\nonumber\\
  &\quad-h\sum_{m=-s}^{s}\hat{\bm{c}}^\dagger_m\sigma^x\hat{\bm{c}}_m\ ,
  \label{H_F_simple_c}
\end{align}
where $V^{\mathcal{F},i}$ $(i=0,1)$ are obtained by replacing 
$\mathcal{F}_{\mathrm{orig}}$ in \eqref{V1234i} with $\mathcal{F}$:
\begin{align} V_{m_1,m_2,m_3,m_4}^{\mathcal{F},i}&=\sum_{k=0}^{2s}U_k^i\,(-1)^{2s+m_2+m_4}(2s+1)
 \begin{pmatrix} s & k & s\\ -m_1 & m_1-m_4 & m_4\end{pmatrix}
 \begin{pmatrix} s & k & s\\ -m_2 & m_2-m_3 & m_3\end{pmatrix}\nonumber\\ &\quad\times\mathcal{F}\big(k(k+1),s\big)\,\delta_{m_1+m_2,m_3+m_4} \ .
 \label{Vi general}
 \end{align}
For
$\mathcal{F}=\mathcal{F}_{\rm orig}$, $\hat{H}_{s,\mathcal{F}}$ reduces to
the simplified Hamiltonian \eqref{H_simple_3j_c}.

Comparing \eqref{V general} with \eqref{V1234}, we see that
$\hat{H}_{\mathcal{F}}$ is identical to an LLL Hamiltonian of the form
\eqref{H_original_psi} with the interaction modified as
\begin{equation}
  U_k\ \to\ U_k\,\frac{\mathcal{F}\big(k(k+1),s\big)}{\mathcal{F}_{\rm orig}\big(k(k+1),s\big)}\ ,
  \label{U_k_modified}
\end{equation}
and the same holds for $\hat{H}_{s,\mathcal{F}}$ in each channel. Hence
$\hat{H}_{\mathcal{F}}$ and $\hat{H}_{s,\mathcal{F}}$ possess the three
symmetries \eqref{Z2 sym}, \eqref{PH sym}, and \eqref{SO(3) sym} for any
$\mathcal{F}$.
Note that after this deformation, $U(\Omega_{ab})$ in \eqref{H_original_psi} includes higer derivative terms. 
In the following, we keep
$R^2=N=2s+1$, as in the numerical calculations.

\subsection{Vector models} \label{sec:F_vector}
Let $L_i$ $(i=1,2,3)$ be the generators of the spin-$s$ representation of
$SU(2)$ acting on the $N$-dimensional space spanned by
\begin{equation}
  \ket{sm}\quad (m=-s,-s+1,\ldots,s)
\end{equation}
with
\begin{align}
  &L_\pm\ket{sm}=\sqrt{(s\mp m)(s\pm m+1)}\,\ket{s\,m\pm1}\ , \qquad
  L_3\ket{sm}=m\ket{sm}\ ,
\end{align}
where $L_\pm=L_1\pm iL_2$. These are the matrix representation of the
angular momentum \eqref{L} on the LLL. We define two fermionic column
vectors $\hat{w}_\alpha$ $(\alpha=\uparrow,\downarrow)$ by
\begin{equation}
  \hat{w}_\alpha=\sum_{m=-s}^{s}\hat{c}^\dagger_{m,\alpha}\ket{sm}\ ,
  \qquad
  \hat{w}^\dagger_\alpha=\sum_{m=-s}^{s}\hat{c}_{m,\alpha}\bra{sm}\ .
  \label{fa}
\end{equation}
The $i$-th component of $\hat{w}_\alpha$ is the fermion operator
$\hat{c}^\dagger_{i-s-1,\alpha}$. Note that the space on which
$\hat{c}^\dagger_{m,\alpha}$ acts is not the space spanned by $\ket{sm}$. We
also define $N\times N$ matrices whose entries are fermion bilinears,
\begin{equation}
  \hat{W}^{(a)}=\hat{w}\,\sigma^a\hat{w}^\dagger
  \equiv\sum_{\alpha_1,\alpha_2}\hat{w}_{\alpha_1}\sigma^a_{\alpha_1\alpha_2}\hat{w}^\dagger_{\alpha_2}
  =\sum_{m_1,m_2}\big(\hat{\bm{c}}^\dagger_{m_1}\sigma^a\hat{\bm{c}}_{m_2}\big)\ket{sm_1}\bra{sm_2}
  \quad (a=0,x,y,z)\ ,
  \label{Fa}
\end{equation}
and the linear maps $L^{\rm ad}_i X=[L_i,X]$ on $N\times N$ matrices $X$.
They satisfy
\begin{equation}
  \Tr\big(XL_i^{\rm ad}Y\big)=-\Tr\big((L_i^{\rm ad}X)Y\big)\ , \qquad
  \big[L^{\rm ad}_i,(L^{\rm ad})^2\big]=0\ ,
  \label{Lad_properties}
\end{equation}
where $\Tr$ is the trace over $N\times N$ matrices and
$(L^{\rm ad})^2\equiv L^{\rm ad}_iL^{\rm ad}_i$. 
More generally, 
for a function $\kappa(x)$, we have
\begin{equation}
  \Tr\Big(X\,\kappa\big((L^{\rm ad})^2\big)Y\Big)=\Tr\Big(\big(\kappa\big((L^{\rm ad})^2\big)X\big)\,Y\Big)\ , \;\;\;
  [L_i^{\rm ad}, \kappa\big((L^{\rm ad})^2\big)]=0 \ .
  \label{kappa relation}
\end{equation}

We introduce the fuzzy spherical harmonics $\hat{Y}_{km}$
$(k=0,1,\ldots,2s,\ m=-k,\ldots,k)$ \cite{Hoppe:1988gk,Grosse:1995jt,Dasgupta:2002hx,Ishiki:2006yr,Ishii:2008tm,Ishii:2008ib},
\begin{equation}
  \hat{Y}_{km}=\sqrt{2s+1}\sum_{n,n'=-s}^{s}(-1)^{-s+n'}C^{km}_{sns(-n')}\ket{sn}\bra{sn'}\ ,
\end{equation}
where $C^{km}_{sns(-n')}$ are Clebsch--Gordan coefficients. They serve as
a basis of $N\times N$ Hermitian matrices and satisfy
\begin{align}
  &\big[L_\pm,\hat{Y}_{km}\big]=\sqrt{(k\mp m)(k\pm m+1)}\,\hat{Y}_{k(m\pm1)}\ , \qquad
  \big[L_3,\hat{Y}_{km}\big]=m\,\hat{Y}_{km}\ ,
  \label{commutator between L and Y}\\
  &\big(\hat{Y}_{km}\big)^\dagger=(-1)^m\hat{Y}_{k(-m)}\ ,
  \label{conjugate of Y}\\
  &\frac{1}{2s+1}\Tr\Big(\big(\hat{Y}_{km}\big)^\dagger\hat{Y}_{k'm'}\Big)=\delta_{kk'}\delta_{mm'}\ ,
  \label{Tr YY}
\end{align}
and, in particular,
\begin{equation}
  (L^{\rm ad})^2\hat{Y}_{km}
=k(k+1)\,\hat{Y}_{km}\ , \;\;\;
  \kappa\big((L^{\rm ad})^2\big) 
  \hat{Y}_{km} = \kappa(k(k+1)) \hat{Y}_{km} \ .
  \label{LLY}
\end{equation}
Using the orthogonality of Clebsch--Gordan coefficients, $\hat{W}^{(a)}$
is expanded as
\begin{equation}
  \hat{W}^{(a)}=\frac{1}{\sqrt{2s+1}}\sum_{k=0}^{2s}\sum_{m=-k}^{k}\hat{W}^{(a)}_{km}\hat{Y}_{km}\ ,
  \qquad
  \hat{W}^{(a)}_{km}=\sum_{m_1,m_4}C^{km}_{sm_1s(-m_4)}(-1)^{s-m_4}
  \big(\hat{\bm{c}}^\dagger_{m_1}\sigma^a\hat{\bm{c}}_{m_4}\big)\ .
  \label{Fakm}
\end{equation}
Then, \eqref{conjugate of Y}, \eqref{Tr YY}, and \eqref{LLY} give, for any
function $\kappa$ and $a,b=0,x,y,z$,
\begin{align}
  \Tr\Big(\hat{W}^{(a)}\kappa\big((L^{\rm ad})^2\big)\hat{W}^{(b)}\Big)
  &=\sum_{k=0}^{2s}\sum_{m=-k}^{k}\kappa\big(k(k+1)\big)(-1)^m\,
  \hat{W}^{(a)}_{km}\hat{W}^{(b)}_{k(-m)}\nonumber\\
  &=\sum_{m_{1,2,3,4}}\sum_{k=0}^{2s}(2k+1)\,\kappa\big(k(k+1)\big)(-1)^{2s+m_2+m_4}\nonumber\\
  &\qquad\times
  \begin{pmatrix} s & k & s\\ -m_1 & m_1-m_4 & m_4\end{pmatrix}
  \begin{pmatrix} s & k & s\\ -m_2 & m_2-m_3 & m_3\end{pmatrix}
  \delta_{m_1+m_2,m_3+m_4}\nonumber\\
  &\qquad\times
  \big(\hat{\bm{c}}^\dagger_{m_1}\sigma^a\hat{\bm{c}}_{m_4}\big)
  \big(\hat{\bm{c}}^\dagger_{m_2}\sigma^b\hat{\bm{c}}_{m_3}\big)\ ,
  \label{trace_lemma}
\end{align}
where, in the second equality, we have written the Clebsch--Gordan
coefficients in terms of $3j$ symbols\footnote{The Clebsch--Gordan
coefficients can be written in terms of $3j$ symbols as follows. 
\begin{equation}
    C_{j_1 m_1 j_2 m_2}^{j_3 m_3} = 
    (-1)^{j_1 - j_2 + m_3} \sqrt{2 j_3 + 1} 
    \begin{pmatrix} j_1 & j_2 & j_3 \\ m_1 & m_2 & -m_3 \end{pmatrix} \ . \nonumber
\end{equation}
}. Comparing \eqref{trace_lemma} with
\eqref{V general} and \eqref{U_k}, and using $R^2=N$, we find
\begin{equation}
  \frac12\sum_{m_{1,2,3,4}}V^{\mathcal{F}}_{m_1,m_2,m_3,m_4}
  \big(\hat{\bm{c}}^\dagger_{m_1}\sigma^a\hat{\bm{c}}_{m_4}\big)
  \big(\hat{\bm{c}}^\dagger_{m_2}\sigma^a\hat{\bm{c}}_{m_3}\big)
  =\frac12\Tr\Big(\hat{W}^{(a)}\,\mathcal{U}\big((L^{\rm ad})^2\big)\,
  \mathcal{F}\big((L^{\rm ad})^2,s\big)\,\hat{W}^{(a)}\Big)\ ,
  \label{V_as_trace}
\end{equation}
where
\begin{equation}
  \mathcal{U}(x)=g_0-\frac{g_1}{N}\,x\ ,
  \label{U_operator}
\end{equation}
so that $U_k=\frac{2k+1}{N}\mathcal{U}\big(k(k+1)\big)$. 
Using \eqref{kappa relation}
yields
\begin{align}
&\Tr\Big(\hat{W}^{(a)}\,\mathcal{U}\big((L^{\rm ad})^2\big)\,
  \mathcal{F}\big((L^{\rm ad})^2,s\big)\,\hat{W}^{(a)}\Big)
  \nonumber\\
&=\Tr\Big(\Big(\sqrt{\mathcal{F}\big((L^{\rm ad})^2,s\big)}\hat{W}^{(a)}\Big)\,\mathcal{U}\big((L^{\rm ad})^2\big)\,
  \sqrt{\mathcal{F}\big((L^{\rm ad})^2,s\big)}\,\hat{W}^{(a)}\Big)
  \ ,
  \label{sqrt F}
\end{align}
where $\sqrt{\mathcal{F}}\equiv\sqrt{\mathcal{F}\big((L^{\rm ad})^2,s\big)}$
acts on the matrix to its right.
Using \eqref{V_as_trace}, \eqref{sqrt F}
and relations
\begin{equation}
  \Tr\big(\hat{w}\sigma^x\hat{w}^\dagger\big)=\sum_m\hat{\bm{c}}^\dagger_m\sigma^x\hat{\bm{c}}_m\ ,
  \qquad
  -\Tr\big(X(L^{\rm ad})^2X\big)=\Tr\big([L_i,X]^2\big)\ ,
  \nonumber
\end{equation}
we write the Hamiltonian
\eqref{H_F_c} in terms of the fermionic vectors as
\begin{align}
  \hat{H}_{\mathcal{F}}
  &=\frac12\Tr\left(\frac{g_1}{N}\Big[L_i,\sqrt{\mathcal{F}}\,\hat{w}\hat{w}^\dagger\Big]^2
  +g_0\Big(\sqrt{\mathcal{F}}\,\hat{w}\hat{w}^\dagger\Big)^2\right)\nonumber\\
  &\quad-\frac12\Tr\left(\frac{g_1}{N}\Big[L_i,\sqrt{\mathcal{F}}\,\hat{w}\sigma^z\hat{w}^\dagger\Big]^2
  +g_0\Big(\sqrt{\mathcal{F}}\,\hat{w}\sigma^z\hat{w}^\dagger\Big)^2\right)\nonumber\\
  &\quad-h\Tr\big(\hat{w}\sigma^x\hat{w}^\dagger\big)\ ,
  \label{vector_model_H'}
\end{align}
where 
$\hat{w}\hat{w}^\dagger=\hat{W}^{(0)}$
and
$\hat{w}\sigma^z\hat{w}^\dagger=\hat{W}^{(z)}$. In the same way,
\eqref{H_F_simple_c} is written as
\begin{equation}
  \hat{H}_{s,\mathcal{F}}
  =\Tr\left(-\frac{g_1}{2N}\Big[L_i,\sqrt{\mathcal{F}}\,\hat{w}\sigma^z\hat{w}^\dagger\Big]^2
  +\frac{g_0}{2}\Big(\sqrt{\mathcal{F}}\,\hat{w}\hat{w}^\dagger\Big)^2
  -h\,\hat{w}\sigma^x\hat{w}^\dagger\right)\ ,
  \label{vector_model_H_F_simple}
\end{equation}
which is obtained from \eqref{vector_model_H'} by dropping the
$g_1$ term in the $a=0$ channel and the $g_0$ term in the $a=z$
channel. The interactions are now expressed through the adjoint action of
$L_i$ on the bilinears $\hat{w}\sigma^a\hat{w}^\dagger$, which transform in
the adjoint representation of $SU(2)$. Note that \eqref{vector_model_H'} and
\eqref{vector_model_H_F_simple} do not refer to the orbital basis
$\ket{sm}$ or to the $3j$ symbols, which entered only through
\eqref{trace_lemma}. They are covariant under the simultaneous
transformation
\begin{equation}
  \hat{w}_\alpha\to U\hat{w}_\alpha\ , \qquad
  L_i\to UL_iU^\dagger\ , \qquad U\in U(N)\ ,
  \label{U(N)_covariance}
\end{equation}
since $\kappa\big((L^{\rm ad})^2\big)$ built from $UL_iU^\dagger$ acts on
$UXU^\dagger$ as $U\big(\kappa((L^{\rm ad})^2)X\big)U^\dagger$. The
$SU(2)$ rotation symmetry \eqref{SO(3) sym} is the subgroup of
\eqref{U(N)_covariance} that leaves $L_i$ invariant up to an $SO(3)$
rotation of the index $i$.

We next move to the path-integral formalism. We replace
$\hat{c}_{m,\alpha}$ and $\hat{w}_\alpha$ by $\tau$-dependent Grassmann
variables $c_{m,\alpha}(\tau)$ and $w_\alpha(\tau)$, and introduce a
Lagrange multiplier $\lambda(\tau)$ to fix the electron number to $N$. The
partition function is
\begin{equation}
  Z=\int\mathcal{D}c\,\mathcal{D}c^*\mathcal{D}\lambda\
  \exp\bigg(-\int d\tau\Big(\sum_{m,\alpha}c^*_{m,\alpha}\partial_\tau c_{m,\alpha}+H_{\mathcal{F}}\Big)
  +i\int d\tau\,\lambda\Big(\sum_{m,\alpha}c^*_{m,\alpha}c_{m,\alpha}-N\Big)\bigg)\ ,
  \label{partition function}
\end{equation}
where $H_{\mathcal{F}}$ is obtained from \eqref{vector_model_H'} by
replacing $\hat{w}$ and $\hat{w}^\dagger$ with $w(\tau)$ and
$w^\dagger(\tau)$.
Using
$\Tr\big(Xw\sigma^aw^\dagger\big)=-w^\dagger X\sigma^aw$ for $a=0, x, z$, we obtain the
vector model
\begin{equation}
  Z=\int\mathcal{D}w\,\mathcal{D}w^*\mathcal{D}\lambda\ e^{-S_{\rm v}}
  \label{Z_vector}
\end{equation}
with
\begin{align}
  S_{\rm v}
  &=\int d\tau\Bigg(w^\dagger\partial_\tau w+h\,w^\dagger\sigma^xw+i\lambda\big(w^\dagger w+N\big)\nonumber\\
  &\qquad\quad+\Tr\bigg(\frac{g_1}{2N}\Big[L_i,\sqrt{\mathcal{F}}\,ww^\dagger\Big]^2
  +\frac{g_0}{2}\Big(\sqrt{\mathcal{F}}\,ww^\dagger\Big)^2\nonumber\\
  &\qquad\qquad\quad-\frac{g_1}{2N}\Big[L_i,\sqrt{\mathcal{F}}\,w\sigma^zw^\dagger\Big]^2
  -\frac{g_0}{2}\Big(\sqrt{\mathcal{F}}\,w\sigma^zw^\dagger\Big)^2\bigg)\Bigg)\ ,
  \label{S_vector}
\end{align}
which is described by two $N$-dimensional Grassmann column vectors
$w_\alpha(\tau)$ and the scalar $\lambda(\tau)$. The vector model for
$\hat{H}_{s,\mathcal{F}}$ is obtained by replacing the last two lines of
\eqref{S_vector} with the interaction terms of
\eqref{vector_model_H_F_simple}:
\begin{align}
  S_{s,{\rm v}}
  &=\int d\tau\Bigg(w^\dagger\partial_\tau w+h\,w^\dagger\sigma^xw+i\lambda\big(w^\dagger w+N\big)\nonumber\\
  &\qquad\quad
  +\Tr\bigg(-\frac{g_1}{2N}\Big[L_i,\sqrt{\mathcal{F}}\,w\sigma^zw^\dagger\Big]^2
  +\frac{g_0}{2}\Big(\sqrt{\mathcal{F}}\,ww^\dagger\Big)^2\bigg)\Bigg)\ .
  \label{S_simple_vector}
\end{align}

\subsection{Matrix--vector models} \label{sec:F_matrix_vector}

The interaction terms in \eqref{S_vector} are quadratic in the bilinears
$ww^\dagger$ and $w\sigma^zw^\dagger$. We make them linear by a
Hubbard--Stratonovich transformation with eight bosonic $N\times N$
Hermitian matrices $\phi^{(a)}_\mu(\tau)$ $(a=0,z,\ \mu=0,1,2,3)$:
\begin{equation}
  Z=\int\mathcal{D}w\,\mathcal{D}w^*\mathcal{D}\lambda\,\mathcal{D}\phi\ e^{-S'_{\rm mv}}
\end{equation}
with
\begin{align}
  S'_{\rm mv}
  &=\int d\tau\Bigg(w^\dagger\partial_\tau w+h\,w^\dagger\sigma^xw+i\lambda\big(w^\dagger w+N\big)
  +\Tr\bigg(\frac12\big(\phi^{(0)}_\mu\big)^2+\frac12\big(\phi^{(z)}_\mu\big)^2\nonumber\\
  &\qquad\quad
  +i\sqrt{\frac{g_1}{N}}\,\phi^{(0)}_i\Big[L_i,\sqrt{\mathcal{F}}\,ww^\dagger\Big]
  +i\sqrt{g_0}\,\phi^{(0)}_0\sqrt{\mathcal{F}}\,ww^\dagger\nonumber\\
  &\qquad\quad
  +\sqrt{\frac{g_1}{N}}\,\phi^{(z)}_i\Big[L_i,\sqrt{\mathcal{F}}\,w\sigma^zw^\dagger\Big]
  +\sqrt{g_0}\,\phi^{(z)}_0\sqrt{\mathcal{F}}\,w\sigma^zw^\dagger\bigg)\Bigg)\ ,
  \label{S_mv_phi}
\end{align}
where the sum over $\mu$ runs from $0$ to $3$ and the sum over $i$ from $1$
to $3$. Integrating out $\phi$ reproduces \eqref{S_vector}. Note that
$[L_i,X]$ is anti-Hermitian for a Hermitian $X$, so that the terms in
\eqref{S_vector} of either sign are generated with the appropriate
factors of $i$.\footnote{Here we assume $g_0,g_1>0$. For $g_1<0$, which
occurs at criticality for the Gaussian deformation of the original model at large $\xi$
(section \ref{sec:gaussian_original}), \eqref{S_vector} is unchanged, and the
formulas below hold with $\sqrt{g_1/N}$ replaced by $i\sqrt{|g_1|/N}$.}
Using
\eqref{Lad_properties} and \eqref{kappa relation}, we move $[L_i,\,\cdot\,]$ and $\sqrt{\mathcal{F}}$
onto $\phi$, and change variables as
\begin{equation}
A_\mu=\sqrt{\mathcal{F}}\,\phi^{(0)}_\mu\ , \qquad
B_\mu=\sqrt{\mathcal{F}}\,\phi^{(z)}_\mu\ .
\end{equation}
The Jacobian is a constant. Using
$\Tr\big(Xw\sigma^aw^\dagger\big)=-w^\dagger X\sigma^aw$  for $a=0, x, z$, 
we obtain
the matrix--vector model
\begin{equation}
  Z=\int\mathcal{D}A\,\mathcal{D}B\,\mathcal{D}w\,\mathcal{D}w^*\mathcal{D}\lambda\ e^{-S_{\rm mv}}
  \label{Z_mv}
\end{equation}
with
\begin{align}
  S_{\rm mv}
  &=\int d\tau\Bigg(\Tr\Big(\frac12A_\mu\,\mathcal{F}^{-1}A_\mu
  +\frac12B_\mu\,\mathcal{F}^{-1}B_\mu\Big)
  +w^\dagger\partial_\tau w+h\,w^\dagger\sigma^xw+i\lambda\big(w^\dagger w+N\big)\nonumber\\
  &\qquad\quad
  +i\sqrt{\frac{g_1}{N}}\,w^\dagger[L_i,A_i]w-i\sqrt{g_0}\,w^\dagger A_0w
  +\sqrt{\frac{g_1}{N}}\,w^\dagger[L_i,B_i]\sigma^zw-\sqrt{g_0}\,w^\dagger B_0\sigma^zw\Bigg)\ ,
  \label{S_mv}
\end{align}
where $\mathcal{F}^{-1}\equiv\mathcal{F}\big((L^{\rm ad})^2,s\big)^{-1}$
acts on the matrix to its right. The model is described by eight bosonic
$N\times N$ Hermitian matrices $A_\mu(\tau)$ and $B_\mu(\tau)$, two
$N$-dimensional Grassmann column vectors $w_\alpha(\tau)$, and the scalar
$\lambda(\tau)$. The bosonic matrices couple to the fermionic vectors
through $[L_i,A_i]$ and $[L_i,B_i]$, that is, as fluctuations around the
fuzzy-sphere background $L_i$. They have no time-derivative term and are
auxiliary fields. Expanding $\Phi=\sum_{k,m}\Phi_{km}\hat{Y}_{km}$ for
$\Phi=A_\mu,B_\mu$, the propagator is proportional to
$\mathcal{F}\big(k(k+1),s\big)$. The action \eqref{S_mv} is covariant under
\eqref{U(N)_covariance} together with $A_\mu\to UA_\mu U^\dagger$ and
$B_\mu\to UB_\mu U^\dagger$. The fuzzy spherical harmonics are merely a
convenient eigenbasis of $(L^{\rm ad})^2$, which is not needed in the
formulation.

For the simplified model, the same procedure with four bosonic matrices
$C_\mu$ gives
\begin{align}
  S_{s,{\rm mv}}
  &=\int d\tau\Bigg(\Tr\Big(\frac12C_\mu\,\mathcal{F}^{-1}C_\mu\Big)
  +w^\dagger\partial_\tau w+h\,w^\dagger\sigma^xw+i\lambda\big(w^\dagger w+N\big)\nonumber\\
  &\qquad\quad
  -i\sqrt{g_0}\,w^\dagger C_0w+\sqrt{\frac{g_1}{N}}\,w^\dagger[L_i,C_i]\sigma^zw\Bigg)\ .
  \label{S_simple_mv}
\end{align}

\subsection{Reduction to scalar matrices} \label{sec:F_eta}

The spatial matrices $A_i$ couple to the fermions only through the
divergence $[L_i,A_i]$. 
This implies that the bosonic sector
is effectively described by one matrix.
Here we make this explicit. 
Any three matrices $A_i$
can be decomposed orthogonally as
\begin{equation}
  A_i=-i\big[L_i,\eta_A\big]+A^{\perp}_i\ , \qquad
  \big[L_i,A^{\perp}_i\big]=0\ , \qquad \Tr\,\eta_A=0\ ,
  \label{eta_decomposition}
\end{equation}
where $\eta_A$ is a traceless Hermitian matrix. Since $\mathcal{F}^{-1}$
commutes with $L^{\rm ad}_i$, the quadratic term also decomposes,
\begin{align}
  \frac12\Tr\big(A_i\,\mathcal{F}^{-1}A_i\big)
  =-\frac12\Tr\Big(\big[L_i,\eta_A\big]\,\mathcal{F}^{-1}\big[L_i,\eta_A\big]\Big)
  +\frac12\Tr\big(A^{\perp}_i\,\mathcal{F}^{-1}A^{\perp}_i\big)\ ,
\end{align}
and the divergence is expressed in 
terms of $\eta_A$ as
\begin{align}
  i\big[L_i,A_i\big]=(L^{\rm ad})^2\eta_A\ .\nonumber
\end{align}
The transverse part $A^{\perp}_i$ decouples and gives a free Gaussian
integral, and the Jacobian of the change of variables is a constant. The
same applies to $B_i$. We thus obtain a matrix--vector model with four
bosonic matrices $A_0$, $\eta_A$, $B_0$, and $\eta_B$:
\begin{align}
  S^{\eta}_{\rm mv}
  =\int d\tau\Bigg(&\Tr\Big(\frac12A_0\,\mathcal{F}^{-1}A_0+\frac12B_0\,\mathcal{F}^{-1}B_0\nonumber\\
  &\qquad-\frac12\big[L_i,\eta_A\big]\,\mathcal{F}^{-1}\big[L_i,\eta_A\big]
  -\frac12\big[L_i,\eta_B\big]\,\mathcal{F}^{-1}\big[L_i,\eta_B\big]\Big)\nonumber\\
  &+w^\dagger\partial_\tau w+h\,w^\dagger\sigma^xw+i\lambda\big(w^\dagger w+N\big)\nonumber\\
  &-i\sqrt{g_0}\,w^\dagger A_0w+\sqrt{\frac{g_1}{N}}\,w^\dagger\Big((L^{\rm ad})^2\eta_A\Big)w\nonumber\\
  &-\sqrt{g_0}\,w^\dagger B_0\sigma^zw
  -i\sqrt{\frac{g_1}{N}}\,w^\dagger\Big((L^{\rm ad})^2\eta_B\Big)\sigma^zw\Bigg)\ .
  \label{S_mv_eta}
\end{align}
The quadratic term of $\eta_A$ and $\eta_B$ has the same form as the
conventional kinetic term $-\Tr[L_i,\Phi]^2$ of fuzzy-sphere field theories
such as \eqref{fuzzy MQM}, dressed by $\mathcal{F}^{-1}$, but it carries no
time derivative. The formulations \eqref{S_mv} and \eqref{S_mv_eta} are equivalent. The former, in which the spatial matrices form a vector $A_i$, is suggestive of
fluctuations $X_i=L_i+A_i$ around a fuzzy-sphere background in matrix
models, while the latter contains the minimal number of bosonic matrices.

For the simplified model, $C_i$ is reduced in the same way
to a single traceless Hermitian matrix $\eta_C$:
\begin{align}
  S^{\eta}_{s,{\rm mv}}
  =\int d\tau\Bigg(&\Tr\Big(\frac12C_0\,\mathcal{F}^{-1}C_0
  -\frac12\big[L_i,\eta_C\big]\,\mathcal{F}^{-1}\big[L_i,\eta_C\big]\Big)\nonumber\\
  &+w^\dagger\partial_\tau w+h\,w^\dagger\sigma^xw+i\lambda\big(w^\dagger w+N\big)\nonumber\\
  &-i\sqrt{g_0}\,w^\dagger C_0w
  -i\sqrt{\frac{g_1}{N}}\,w^\dagger\Big((L^{\rm ad})^2\eta_C\Big)\sigma^zw\Bigg)\ .
  \label{S_simple_mv_eta}
\end{align}

\subsection{Matrix models} \label{sec:F_matrix}

Since \eqref{S_mv} is quadratic in $w$ and $w^*$, we can integrate them out.
We rewrite $S_{\rm mv}$ as
\begin{equation}
  S_{\rm mv}=\int d\tau\left(\Tr\Big(\frac12A_\mu\,\mathcal{F}^{-1}A_\mu
  +\frac12B_\mu\,\mathcal{F}^{-1}B_\mu\Big)+iN\lambda+w^\dagger\mathcal{M}(A,B,\lambda)\,w\right)
\end{equation}
with
\begin{equation}
  \mathcal{M}(A,B,\lambda)=\partial_\tau+h\sigma^x+i\lambda
  +i\sqrt{\frac{g_1}{N}}\,[L_i,A_i]-i\sqrt{g_0}\,A_0
  +\sqrt{\frac{g_1}{N}}\,[L_i,B_i]\sigma^z-\sqrt{g_0}\,B_0\sigma^z\ .
  \label{M_F}
\end{equation}
Then,
the integration over $w$ and $w^*$ gives $\det\mathcal{M}(A,B,\lambda)$,
which is complex in general. We thus obtain the matrix model
\begin{equation}
Z=\int\mathcal{D}A\,\mathcal{D}B\,\mathcal{D}\lambda\ e^{-S_{\rm m}}
  \label{Z_m}
\end{equation}
with
\begin{equation}
  S_{\rm m}=\int d\tau\left(\Tr\Big(\frac12A_\mu\,\mathcal{F}^{-1}A_\mu
+\frac12B_\mu\,\mathcal{F}^{-1}B_\mu\Big)+iN\lambda\right)
  -\int d\tau\,\tr\log\mathcal{M}(A,B,\lambda)\ ,
  \label{S_m}
\end{equation}
where $\tr$ is the trace over the tensor product of $N\times N$ matrices and
$2\times2$ matrices acting on the pseudospin index. In the formulation
\eqref{S_mv_eta}, $\mathcal{M}$ is replaced by
\begin{equation}
  \mathcal{M}^{\eta}=\partial_\tau+h\sigma^x+i\lambda-i\sqrt{g_0}\,A_0
  +\sqrt{\frac{g_1}{N}}\,(L^{\rm ad})^2\eta_A
  -\sqrt{g_0}\,B_0\sigma^z-i\sqrt{\frac{g_1}{N}}\,\Big((L^{\rm ad})^2\eta_B\Big)\sigma^z\ .
\end{equation}
For the simplified model, we obtain
\begin{equation}
  S_{s,{\rm m}}=\int d\tau\left(\Tr\Big(\frac12C_\mu\,\mathcal{F}^{-1}C_\mu\Big)+iN\lambda\right)
  -\int d\tau\,\tr\log\mathcal{M}_s(C,\lambda)\ ,
  \label{S_simple_m}
\end{equation}
\begin{equation}
  \mathcal{M}_s(C,\lambda)=\partial_\tau+h\sigma^x+i\lambda-i\sqrt{g_0}\,C_0
  +\sqrt{\frac{g_1}{N}}\,[L_i,C_i]\sigma^z\ .
  \label{M_F_simple}
\end{equation}
Since
$\big(i\mathcal{M}_s(C,\lambda)\big)^\dagger=\sigma^z\big(i\mathcal{M}_s(C,\lambda)\big)\sigma^z$,
$\det\big(i\mathcal{M}_s(C,\lambda)\big)$ is real for $g_0,g_1>0$. The same holds in the
formulation \eqref{S_simple_mv_eta}.

To summarize, every Hamiltonian in the class \eqref{H_F_c}, including the
original model at $\mathcal{F}=\mathcal{F}_{\rm orig}$, is equivalent to
the matrix--vector model \eqref{Z_mv} with \eqref{S_mv}, or \eqref{S_mv_eta}\footnote{For \eqref{S_mv_eta}, the path integral measure in \eqref{Z_mv} is replaced by $\int\mathcal{D}A_0\,\mathcal{D}\eta_A\,\mathcal{D}w\,\mathcal{D}w^*\mathcal{D}\lambda$.},
and to the matrix model \eqref{Z_m} with \eqref{S_m}. The same holds for the class
\eqref{H_F_simple_c}. All of these models are written in the
basis-independent form described above, in which the sphere is encoded only
in the background $L_i$. The choice of $\mathcal{F}$ enters only through the
quadratic term of the bosonic matrices. For the Gaussian $\mathcal{F}_\xi$ in
\eqref{F Gaussian},
$\mathcal{F}^{-1}=e^{\frac{\xi}{2s}(L^{\rm ad})^2}$, and the propagator of
the bosonic matrices becomes proportional to the heat kernel $e^{-\xi k(k+1)/2s}$. In the
next section, we show that this choice also gives the 3D Ising CFT at
criticality for a range of $\xi$.

\section{Gaussian deformation and the 3D Ising CFT} \label{deformation}

In this section, we study the Hamiltonians in the class of section \ref{sec:F_class}
with the Gaussian function \eqref{F Gaussian}, for both the original and the
simplified Hamiltonians, and show by exact diagonalization that they give the
3D Ising CFT at criticality for a range of $\xi$.

\subsection{Gaussian deformation} \label{sec:gaussian_models}

We consider the replacement
\begin{equation}
  \mathcal{F}_{\rm orig}\big(k(k+1),s\big)
  =(2s+1)\begin{pmatrix} s & k & s\\ -s & 0 & s\end{pmatrix}^2
  \ \to\
  \mathcal{F}_\xi\big(k(k+1),s\big)=\exp\left(-\frac{\xi k(k+1)}{2s}\right)\ ,
  \label{replacement}
\end{equation}
where $\xi>0$ is a deformation parameter. We refer to the resulting
Hamiltonians
\begin{align}
  \hat{H}'&\equiv\hat{H}_{\mathcal{F}_\xi}\ ,
  \label{H_deformed_c}\\
  \hat{H}'_s&\equiv\hat{H}_{s,\mathcal{F}_\xi}\ ,
  \label{H_simple_deformed_c}
\end{align}
defined by \eqref{H_F_c} and \eqref{H_F_simple_c}, as the Gaussian models.
By the results in 
section \ref{matrix-vector model}, they are equivalent to the matrix--vector
models \eqref{S_mv} and \eqref{S_simple_mv}, and to the matrix models
\eqref{S_m} and \eqref{S_simple_m}, with
$\mathcal{F}^{-1}=e^{\frac{\xi}{2s}(L^{\rm ad})^2}$.
These matrix models take simpler forms
than the ones for $\mathcal{F}_{\rm orig}$.

The Gaussian form is motivated by the large-$s$ behavior of
$\mathcal{F}_{\rm orig}$. Applying Stirling's formula to \eqref{F orig}, one
finds, for $k\ll s^{3/4}$,
\begin{equation}
  \mathcal{F}_{\rm orig}\big(k(k+1),s\big)
  =\exp\left(-\frac{k(k+1)}{2s}\right)
  \left[1+O\!\left(\frac{k^2}{s^2},\frac{k^4}{s^3}\right)\right]\ .
  \label{regularization_factor}
\end{equation}
Thus $\mathcal{F}_{\rm orig}$ suppresses the interaction at
$k\gtrsim\sqrt{s}$, and in the relevant range $k=O(\sqrt{s})$\footnote{Note that $F_{\rm orig}(k(k+1),s)$ decreases monotonically with $k$.} it coincides
with $\mathcal{F}_{\xi=1}$ up to $O(1/s)$ corrections. The Gaussian model at
$\xi=1$ is therefore the original model up to corrections that vanish as
$s\to\infty$. For $\xi\neq1$, in contrast, the difference between the two
functions does not vanish in this limit. With $x=k(k+1)/2s$,
\begin{equation}
  \lim_{s\to\infty}\,\max_k\Big|\mathcal{F}_\xi-\mathcal{F}_{\rm orig}\Big|
  =\max_{x\ge0}\Big|e^{-\xi x}-e^{-x}\Big|
  =\xi^{-\frac{1}{\xi-1}}\left|1-\frac{1}{\xi}\right|\ ,
\end{equation}
which is nonzero for $\xi\neq1$. 
Note that the relevant range of $k$ for $\mathcal{F}_\xi$ is
also $k = O(\sqrt{s})$. By \eqref{U_k_modified}, the Gaussian
model for $\xi\neq1$ is an LLL model with the interaction
$U_k\,\mathcal{F}_\xi/\mathcal{F}_{\rm orig}$. This makes the Haldane
pseudopotentials $V_m$ with $m\ge2$ nonvanishing, whereas they vanish for the
original interaction \eqref{U_k}. 

\subsection{Numerical procedure} \label{sec:numerical_procedure}

The Gaussian models possess the Ising $\mathbb{Z}_2$, particle--hole, and
$SO(3)$ symmetries (section \ref{sec:F_class}). We classify the energy
eigenstates with $L_z=0$ into the sectors $(\mathbb{Z}_2,P,\ell)$ defined in
section \ref{sec:review_CFT}, and test whether, in each sector, the spectrum
can be matched to the operators of the 3D Ising CFT up to finite-size
corrections. Our numerical
implementation follows Ref.~\cite{Zhou:2025liv}: our programs are constructed based on  FuzzifiED. We set $R^2=N$ and $h=1$.
Although only one parameter needs to be tuned to reach criticality in
principle, we tune both $g_0$ and $g_1$ to reduce finite-size corrections
from irrelevant operators, as in Ref.~\cite{Zhu:2022gjc}.

We obtain the low-energy eigenstates by exact diagonalization, normalize the
excitation energies by the energy-momentum tensor as described in
section ~\ref{sec:review_CFT}, and order the states by energy in each sector.
Using the conformal-bootstrap data of
Refs.~\cite{Poland:2018epd,Simmons-Duffin:2016wlq}, a part of which includes
the fuzzy-sphere results of Ref.~\cite{Fardelli:2026zas}, we order the
primary and descendant operators of the 3D Ising CFT in the same way. We then
match the numerical states and the CFT operators of the same rank in each
sector\footnote{Some operators in the same sector have the same $\Delta$.
In such cases, the corresponding individual eigenstates cannot be
distinguished by $\Delta$ and the sector alone, as
in the analysis in Appendix D of Ref.~\cite{Zhu:2022gjc}.} and compare their
conformal dimensions. The parameters $g_0$ and $g_1$ are tuned by an
iteration of exact diagonalization and Nelder--Mead optimization, which makes
the differences between the dimensions of the primary operators and their descendants  with $\Delta\le3$ as close as possible to integers. The tuning
procedure is described in Appendix~\ref{g0_g1_tuning}.

We focus on the operators with $\Delta<7$ and $\ell\le4$ in the 3D Ising CFT. 
There are 73 such operators, each represented by
its $L_z=0$ component, of which 14 are primary.
In our numerical results, because of finite-size corrections, some operators
with reference dimensions slightly below 7 have numerical dimensions
slightly above 7, and conversely. 
It turns out that every operator with $\Delta<7$ and $\ell\le4$ in the reference data can be matched to a numerical eigenstate up to $N=16$. 
We quantify the agreement by the mean and
maximum of the relative errors
$|\Delta_{\rm num}-\Delta_{\rm ref}|/\Delta_{\rm ref}$ over these 73
operators.\footnote{For operators identified by the conformal bootstrap, the
reference values are taken from
Refs.~\cite{Poland:2018epd,Simmons-Duffin:2016wlq}. For the
parity-odd primary operator $\sigma^{P-}_{\mu_1\mu_2\mu_3\mu_4}$, which is
not identified in those bootstrap results, we use the value $6.53(5)$ from
Ref.~\cite{Fardelli:2026zas}. This is the only such operator in the range
$\Delta<7$, $\ell\le4$ considered here. \label{note_known_data}} 
Finite-size corrections can produce eigenstates with numerical conformal dimensions below 7 whose reference dimensions are slightly above 7. We confirm that each of such states can also be identified with an operator in the reference data at least at $N=16$. We refer to such states as ``extra states''. In Appendix \ref{sec:n_dependence_extra_states}, we examine extra states. 
We perform
the analysis at $N = 16$ of the Gaussian models with $\xi=0.8,0.9,\ldots,2.5$ for the original model
and of those with $\xi=0.7,0.8,\ldots,1.7$ for the simplified model, as well as
of the original and simplified models
themselves\footnote{The tuning procedure for the
original Hamiltonian differs from that of Ref.~\cite{Zhu:2022gjc}.
Accordingly, the conformal dimensions obtained here differ slightly from
those reported there. See Appendix~\ref{g0_g1_tuning}.}. We also study the $N$ dependence for $8\le N\le16$. We present the numerical results for the original model and its Gaussian models in section \ref{sec:gaussian_original} and for the simplified model and its Gaussian models in section \ref{sec:gaussian_simple}. 

\subsection{Original model} \label{sec:gaussian_original}

\begin{figure}[tb]
  \centering
  \includegraphics[width=0.48\textwidth]{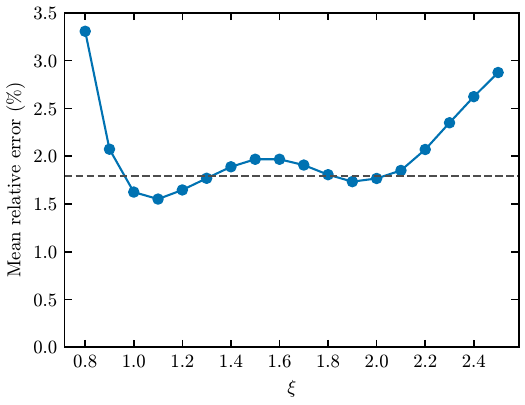}
  \hfill
  \includegraphics[width=0.48\textwidth]{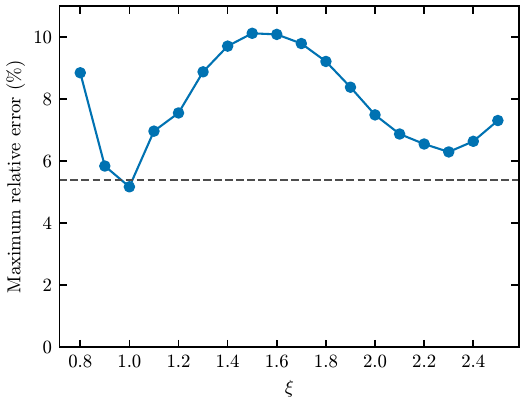}
  \caption{Dependence on $\xi$ of the mean (left) and maximum (right) relative
  errors between the conformal dimensions obtained from $N=16$ exact
  diagonalization of the Gaussian models \eqref{H_deformed_c} and the reference
  values \cite{Poland:2018epd,Simmons-Duffin:2016wlq,Fardelli:2026zas}, for
  the 73 operators with $\Delta<7$ and $\ell\le4$. The dashed lines indicate
  the corresponding results for the original model. 
  }
  \label{fig:xi_dependence_mean_max_original}
\end{figure}

\begin{figure}[tbp]
      \centering
        \includegraphics[width=0.48\linewidth]{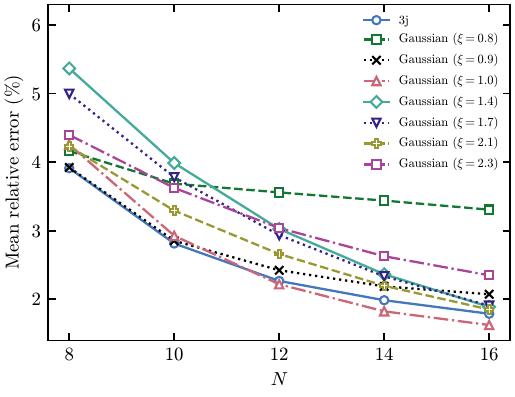}\hfill
        \includegraphics[width=0.48\linewidth]{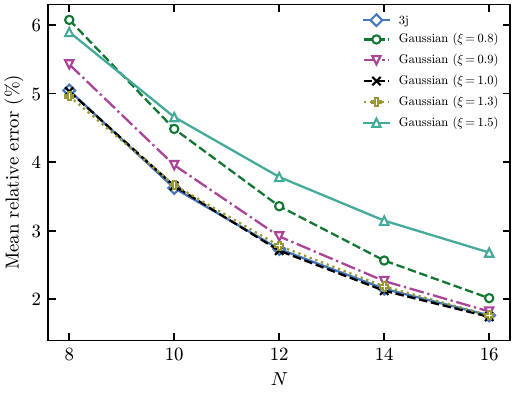}
      \par\vspace{3mm}
        \includegraphics[width=0.48\linewidth]{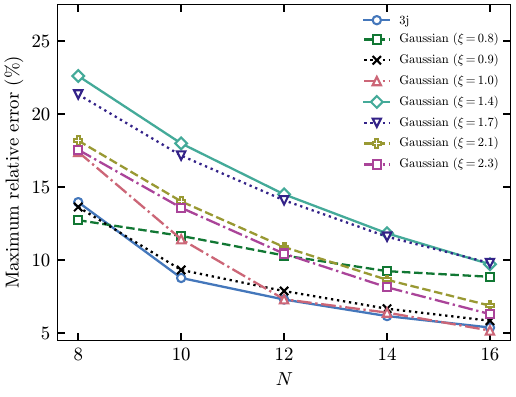}\hfill
        \includegraphics[width=0.48\linewidth]{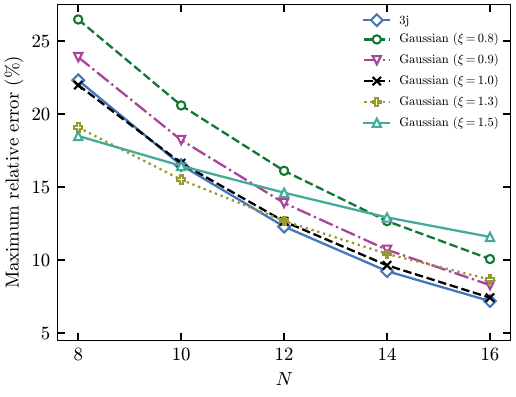}
      \caption{System-size dependence of the mean (upper two panels) and maximum (lower two panels) relative errors in the conformal dimensions of operators with $\Delta < 7$ and $\ell\le4$ for each model. The left panels show the results for the original model \eqref{H_original_c} and its Gaussian models \eqref{H_deformed_c}, while the right panels show the results for the simplified model \eqref{H_simple_3j_c} and its Gaussian models \eqref{H_simple_deformed_c}. ``3j'' stands for
      the orignal model or the simplified model.}
      \label{fig:nm_dependence_mean_max}
    \end{figure}

Fig.~\ref{fig:xi_dependence_mean_max_original} shows the mean and
maximum relative errors as functions of $\xi$ at $N=16$. For $1.0\le\xi\le2.1$, the
mean error is at most $2.0\%$ and the maximum error is at most
$10.1\%$, comparable to those of the original model (mean $1.8\%$ and maximum $5.4\%$). 
The mean error for $\xi=0.8, 0.9, 2.2, 2.3, 2.4, 2.5$ exceeds $2.0\%$, so that 
we exclude the Gaussian models with these values of $\xi$ from the models 
that serve as a regularization of the 3D Ising CFT.
The numbers of operators with relative errors
below $3\%$ for 
$\xi=1.0,1.1,1.2,1.3,1.4,1.5,1.6,1.7,1.8,1.9,2.0,2.1$ are 61, 60, 61, 58, 59, 58, 58, 59, 57, 61, 61 and 58, respectively, while that for the original model is 58. 
In Fig.~\ref{fig:nm_dependence_mean_max}, the mean and maximum errors are plotted against $N$ for fixed $\xi$'s.
They decrease as $N$ increases for $1.0 \leq \xi \leq 2.1$.
In appendix~\ref{sec:n_dependence_extra_states}, we see that the number of extra states decreases with $N$ 
for $1.0 \leq \xi \leq 2.1$.
These imply that the relative errors at $1.0 \leq \xi \leq 2.1$ can be attributed 
to finite-size corrections.

\begin{figure}[htbp]
  \centering
  \includegraphics[width=\textwidth]{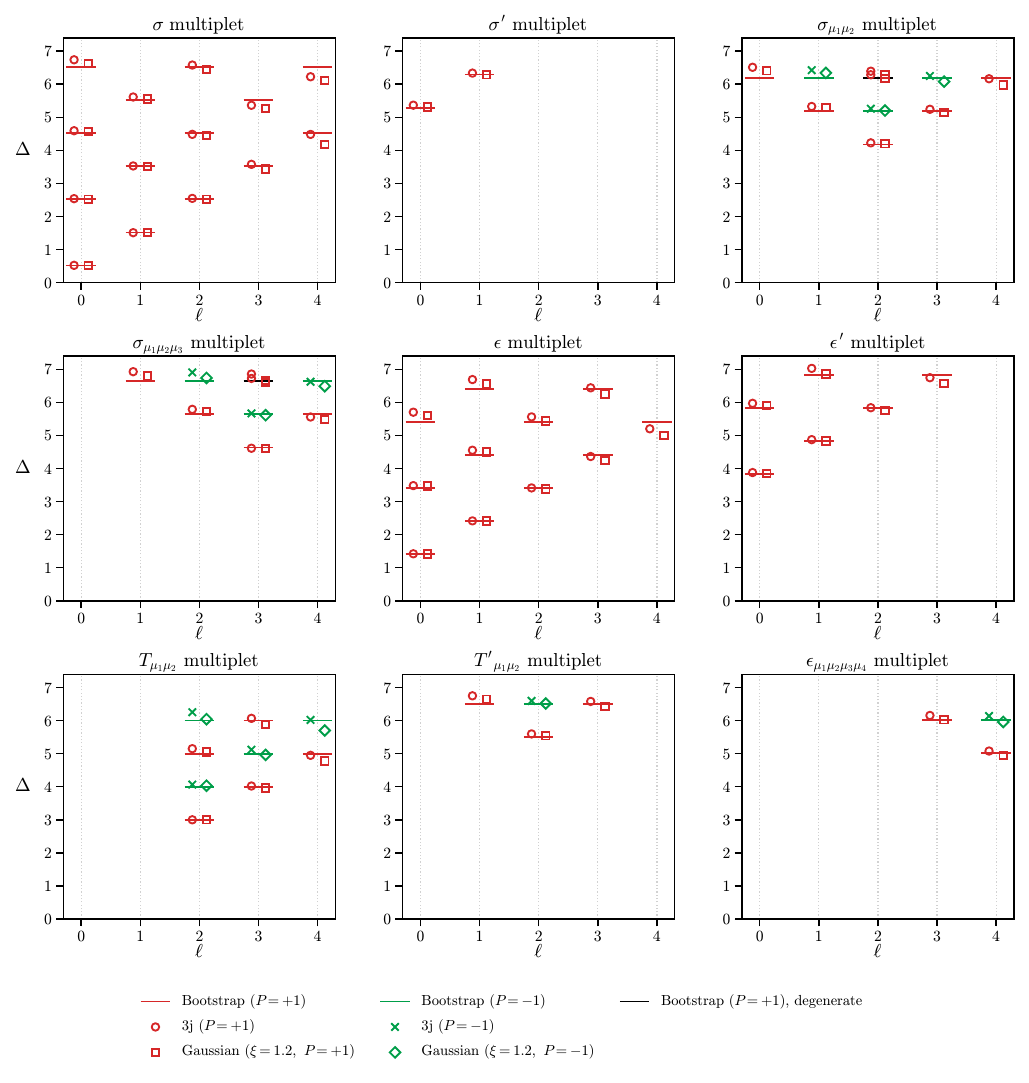}
  \caption{Conformal multiplets of primary operators that have descendants
  within $\Delta<7$ and $\ell\le4$ are shown for the original model and
  its Gaussian model at $\xi=1.2$.
   Each panel shows the spin $\ell$ on the
  horizontal axis and the conformal dimension $\Delta$ on the vertical axis.
  The conformal-bootstrap results of
  Refs.~\cite{Poland:2018epd,Simmons-Duffin:2016wlq} are shown
  by horizontal lines: 
  red lines denote parity-even non-degenerate operators,
  green lines denote parity-odd non-degenerate operators, and black lines
  denote parity-even twofold degenerate operators. For the original model (3j)
  \eqref{H_original_c}, the $N=16$ exact-diagonalization results are indicated
  by red circles (parity even) and green crosses (parity odd). For the
  Gaussian model \eqref{H_deformed_c} at $\xi=1.2$, they are indicated by red
  squares (parity even) and green diamonds (parity odd).}
  \label{fig:ising_multiplets_original_xi1.2}
\end{figure}

Table~\ref{table:primary_dimensions_not_simplified} lists 
the conformal dimensions of the
14 primary operators for the original model and its Gaussian models with 
$1.0 \leq \xi \leq 2.1$.
For instance, among the 14 primary operators, the mean and maximum relative
errors for the Gaussian model at $\xi=1.2$ are $0.9\%$ and $4.0\%$, respectively, and those for the original model are $1.0\%$ and $1.6\%$, respectively.  
Fig.~\ref{fig:ising_multiplets_original_xi1.2} shows the
primary and descendant spectra organized into conformal multiplets for
$\xi=1.2$.

For $\xi=0.9, 2.2, 2.3, 2.4, 2.5$, the mean and maximum errors still 
decrease with $N$
and the number of extra states at $N=16$ remains small. 
The deviations at these excluded values of  $\xi$ may therefore be, at least partly,
finite-size effects, and the range of $\xi$ for which the model approaches
 the 3D Ising CFT as $s\to\infty$ may be wider than that found here.
At $\xi=0.8$, the mean and maximum relative errors depend only weakly
on $N$, in
contrast to the other values of $\xi$, for which the error decreases
steadily with $N$. We therefore cannot conclude that the model at $\xi=0.8$
approaches the 3D Ising CFT within the accessible system sizes.

These results show that, for $1.0\le\xi\le2.1$, the low-energy states of
the Gaussian model at criticality are in one-to-one correspondence with the
operators of the 3D Ising CFT, up to finite-size corrections. The Gaussian
model therefore provides another regularization of the 3D Ising CFT in this
range.

Combined with the results in section \ref{matrix-vector model}, the 3D Ising CFT
 is realized by the matrix--vector model \eqref{Z_mv} with \eqref{S_mv}, or
 \eqref{S_mv_eta}, and by the matrix model \eqref{Z_m} with \eqref{S_m}, where
 $\mathcal{F}^{-1}=e^{\frac{\xi}{2s}(L^{\rm ad})^2}$ with
 $1.0\le\xi\le2.1$ and $g_0$ and $g_1$ tuned to criticality.

\begin{table}[t]
\centering
\caption{Primary operators with $\Delta<7$ and $\ell\le4$, together with
 their conformal dimensions, for the original model. The upper and lower parts list $\mathbb{Z}_2$-odd
 and $\mathbb{Z}_2$-even operators, respectively. ``Bootstrap'' gives the
   conformal-bootstrap results of
   Refs.~\cite{Poland:2018epd,Simmons-Duffin:2016wlq}. 
   The other rows give the $N=16$
   exact-diagonalization results for the original model \eqref{H_original_c}
   (``3j'') and its Gaussian models \eqref{H_deformed_c}
   (``Gaussian''). }
\label{table:primary_dimensions_not_simplified}
\begin{tabular*}{\textwidth}{@{\extracolsep{\fill}}l*{7}{c}@{}}
\toprule
$\mathbb{Z}_2$-odd & $\sigma$ & $\sigma'$ & $\sigma_{\mu_1 \mu_2}$ & $\sigma'_{\mu_1 \mu_2}$ & $\sigma_{\mu_1 \mu_2 \mu_3}$ & $\sigma_{\mu_1 \mu_2 \mu_3 \mu_4}$ & $\sigma^{P-}_{\mu_1 \mu_2 \mu_3 \mu_4}$ \\
\midrule
Bootstrap & 0.518 & 5.291 & 4.180 & 6.987 & 4.638 & 6.113 & $\cdots$ \\
3j & 0.524 & 5.367 & 4.227 & 7.096 & 4.614 & 6.072 & 6.562 \\
Gaussian ($\xi = 1.0$) & 0.522 & 5.355 & 4.220 & 7.075 & 4.611 & 6.040 & 6.537 \\
Gaussian ($\xi = 1.1$) & 0.521 & 5.335 & 4.211 & 7.045 & 4.606 & 5.956 & 6.487 \\
Gaussian ($\xi = 1.2$) & 0.520 & 5.317 & 4.202 & 7.019 & 4.602 & 5.869 & 6.439 \\
Gaussian ($\xi = 1.3$) & 0.518 & 5.304 & 4.195 & 6.998 & 4.599 & 5.808 & 6.404 \\
Gaussian ($\xi = 1.4$) & 0.518 & 5.292 & 4.190 & 6.981 & 4.597 & 5.771 & 6.380 \\
Gaussian ($\xi = 1.5$) & 0.517 & 5.286 & 4.187 & 6.972 & 4.596 & 5.753 & 6.369 \\
Gaussian ($\xi = 1.6$) & 0.518 & 5.283 & 4.186 & 6.970 & 4.596 & 5.754 & 6.369 \\
Gaussian ($\xi = 1.7$) & 0.517 & 5.289 & 4.189 & 6.978 & 4.597 & 5.767 & 6.378 \\
Gaussian ($\xi = 1.8$) & 0.519 & 5.302 & 4.195 & 6.997 & 4.599 & 5.794 & 6.396 \\
Gaussian ($\xi = 1.9$) & 0.519 & 5.324 & 4.205 & 7.027 & 4.601 & 5.833 & 6.422 \\
Gaussian ($\xi = 2.0$) & 0.521 & 5.357 & 4.218 & 7.069 & 4.605 & 5.878 & 6.453 \\
Gaussian ($\xi = 2.1$) & 0.520 & 5.401 & 4.234 & 7.122 & 4.610 & 5.917 & 6.481 \\
\midrule
$\mathbb{Z}_2$-even & $\epsilon$ & $\epsilon'$ & $\epsilon''$ & $T_{\mu_1 \mu_2}$ & $T'_{\mu_1 \mu_2}$ & $\epsilon_{\mu_1 \mu_2 \mu_3 \mu_4}$ & $\epsilon'_{\mu_1 \mu_2 \mu_3 \mu_4}$ \\
\midrule
Bootstrap & 1.413 & 3.830 & 6.896 & 3.000 & 5.509 & 5.023 & 6.421 \\
3j & 1.423 & 3.878 & 6.996 & 3.000 & 5.596 & 5.079 & 6.361 \\
Gaussian ($\xi = 1.0$) & 1.420 & 3.871 & 6.979 & 3.000 & 5.580 & 5.054 & 6.352 \\
Gaussian ($\xi = 1.1$) & 1.416 & 3.857 & 6.952 & 3.000 & 5.560 & 5.004 & 6.339 \\
Gaussian ($\xi = 1.2$) & 1.413 & 3.846 & 6.927 & 3.000 & 5.542 & 4.944 & 6.328 \\
Gaussian ($\xi = 1.3$) & 1.410 & 3.837 & 6.909 & 3.000 & 5.528 & 4.903 & 6.320 \\
Gaussian ($\xi = 1.4$) & 1.408 & 3.830 & 6.892 & 3.000 & 5.517 & 4.879 & 6.314 \\
Gaussian ($\xi = 1.5$) & 1.407 & 3.826 & 6.883 & 3.000 & 5.511 & 4.867 & 6.291 \\
Gaussian ($\xi = 1.6$) & 1.407 & 3.824 & 6.879 & 3.000 & 5.510 & 4.867 & 6.293 \\
Gaussian ($\xi = 1.7$) & 1.408 & 3.828 & 6.887 & 3.000 & 5.515 & 4.876 & 6.311 \\
Gaussian ($\xi = 1.8$) & 1.410 & 3.836 & 6.906 & 3.000 & 5.528 & 4.895 & 6.318 \\
Gaussian ($\xi = 1.9$) & 1.413 & 3.850 & 6.937 & 3.000 & 5.547 & 4.924 & 6.328 \\
Gaussian ($\xi = 2.0$) & 1.417 & 3.870 & 6.983 & 3.000 & 5.575 & 4.958 & 6.340 \\
Gaussian ($\xi = 2.1$) & 1.421 & 3.897 & 7.046 & 3.000 & 5.609 & 4.989 & 6.356 \\
\bottomrule
\end{tabular*}
\end{table}

\clearpage

\clearpage

\subsection{Simplified model} \label{sec:gaussian_simple}
\begin{figure}[tb]
  \centering
  \includegraphics[width=0.48\textwidth]{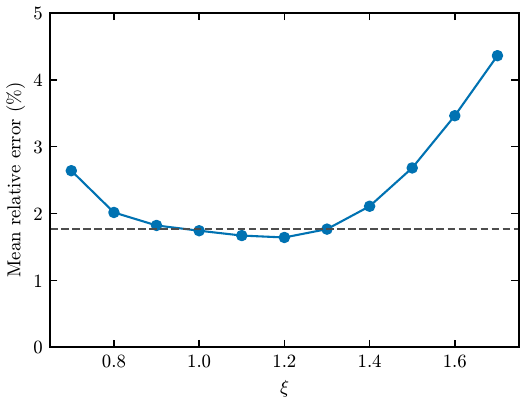}
  \hfill
  \includegraphics[width=0.48\textwidth]{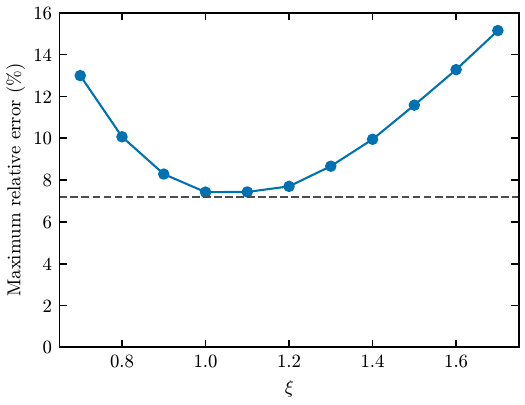}
  \caption{Dependence on $\xi$ of the mean (left) and maximum (right) relative
  errors between the conformal dimensions obtained from $N=16$ exact
  diagonalization of the Gaussian models \eqref{H_simple_deformed_c} and the reference
  values \cite{Poland:2018epd,Simmons-Duffin:2016wlq,Fardelli:2026zas}, for
  the 73 operators with $\Delta<7$ and $\ell\le4$. The dashed lines indicate
  the corresponding results for the simplified model.}
  \label{fig:xi_dependence_mean_max_simplified}
\end{figure}

Fig.~\ref{fig:xi_dependence_mean_max_simplified} shows the mean and
maximum relative errors as functions of $\xi$ at $N=16$. For $0.9\le\xi\le1.3$, the
mean error is at most $1.8\%$ and the maximum error is at most
$8.7\%$, comparable to those of the simplified model (mean $1.8\%$ and maximum $7.2\%$). 
The mean error for $\xi=0.8, 1.4, 1.5, 1.6, 1.7$ exceeds $2.0\%$, so that 
we exclude the Gaussian models with these values of $\xi$ from the models 
that serve as a regularization of the 3D Ising CFT.
The numbers of operators with relative errors
below $3\%$ for 
$\xi=0.9,1.0,1.1,1.2,1.3$ are 60, 60, 61, 61 and 61, respectively, while that for the 
simplified model is 59. 
In Fig.~\ref{fig:nm_dependence_mean_max}, the mean and maximum errors are plotted against $N$ for fixed $\xi$'s.
They decrease as $N$ increases for $0.9 \leq \xi \leq 1.3$.
In appendix~\ref{sec:n_dependence_extra_states}, we see that the number of extra states decreases with $N$ 
for $0.9 \leq \xi \leq 1.3$.
These imply that the relative errors at $0.9 \leq \xi \leq 1.3$ can be attributed 
to finite-size corrections.

\begin{figure}[htbp]
  \centering
  \includegraphics[width=\textwidth]{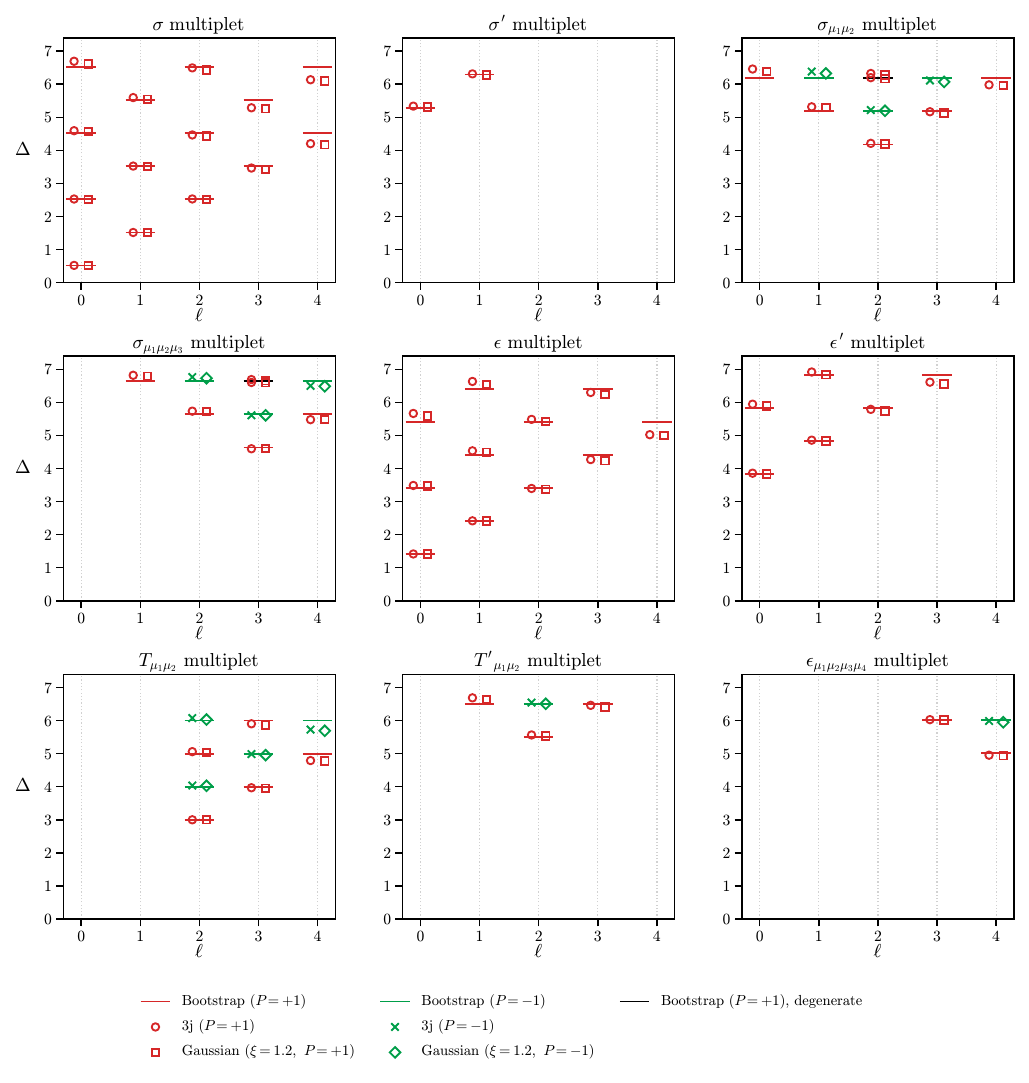}
  \caption{Conformal multiplets of primary operators that have descendants
  within $\Delta<7$ and $\ell\le4$ are shown for the simplified model and
  its Gaussian model at $\xi=1.2$.
   Each panel shows the spin $\ell$ on the
  horizontal axis and the conformal dimension $\Delta$ on the vertical axis.
  The conformal-bootstrap results of
  Refs.~\cite{Poland:2018epd,Simmons-Duffin:2016wlq} are shown
  by horizontal lines: 
  red lines denote parity-even non-degenerate operators,
  green lines denote parity-odd non-degenerate operators, and black lines
  denote parity-even twofold degenerate operators. For the simplified model (3j)
  \eqref{H_simple_3j_c}, the $N=16$ exact-diagonalization results are indicated
  by red circles (parity even) and green crosses (parity odd). For the
  Gaussian model \eqref{H_simple_deformed_c} at $\xi=1.2$, they are indicated by red
  squares (parity even) and green diamonds (parity odd).}
  \label{fig:ising_multiplets_simplified_xi1.2}
\end{figure}

Table~\ref{table:primary_dimensions_simplified} lists 
the conformal dimensions of the
14 primary operators for the simplified model and its Gaussian models with 
$0.9 \leq \xi \leq 1.3$.
For instance, among the 14 primary operators, the mean and maximum relative
errors for the Gaussian model at $\xi=1.2$ are $0.8\%$ and $4.2\%$, respectively, and those for the simplified model are $1.1\%$ and $3.6\%$, respectively.  
Fig.~\ref{fig:ising_multiplets_simplified_xi1.2} shows the
primary and descendant spectra organized into conformal multiplets for
$\xi=1.2$.

For $\xi=0.8, 1.4, 1.5, 1.6, 1.7$, the mean and maximum errors still 
decrease with $N$
and the number of extra states at $N=16$ remains small. 
The deviations at these excluded values of $\xi$ may therefore be, at least partly,
finite-size effects, and the range of $\xi$ for which the model approaches
 the 3D Ising CFT as $s\to\infty$ may be wider than that found here.

These results show that, for $0.9\le\xi\le1.3$, the low-energy states of
the Gaussian model at criticality are in one-to-one correspondence with the
operators of the 3D Ising CFT, up to finite-size corrections. 
Thus, for $0.9\le\xi\le1.3$, the Gaussian deformation of the simplified
model also provides a regularization of the 3D Ising CFT. In contrast to the
original model, the agreement is lost for $\xi\gtrsim1.4$. This shows that
the agreement found above is not automatic for systems of this size, and
that it depends on the choice of $\mathcal{F}$ together with the form of the
interaction.

Combined with the results in section \ref{matrix-vector model}, the 3D Ising CFT
 is realized by the matrix--vector model given by \eqref{S_simple_mv} or
 \eqref{S_simple_mv_eta}, and by the matrix model given by \eqref{S_simple_m}, where
 $\mathcal{F}^{-1}=e^{\frac{\xi}{2s}(L^{\rm ad})^2}$ with
 $0.9\le\xi\le1.3$ and $g_0$ and $g_1$ tuned to criticality.

\begin{table}[tb]
\centering
\caption{Primary operators with $\Delta<7$ and $\ell\le4$, together with
 their conformal dimensions, for the simplified model. The upper and lower parts list $\mathbb{Z}_2$-odd
 and $\mathbb{Z}_2$-even operators, respectively. ``Bootstrap'' gives the
   conformal-bootstrap results of
   Refs.~\cite{Poland:2018epd,Simmons-Duffin:2016wlq}. 
    The other rows give the $N=16$
   exact-diagonalization results for the simplified model \eqref{H_simple_3j_c}
   (``3j'') and its Gaussian models \eqref{H_simple_deformed_c}
   (``Gaussian''). }
\label{table:primary_dimensions_simplified}
\begin{tabular*}{\textwidth}{@{\extracolsep{\fill}}l*{7}{c}@{}}
\toprule
$\mathbb{Z}_2$-odd & $\sigma$ & $\sigma'$ & $\sigma_{\mu_1 \mu_2}$ & $\sigma'_{\mu_1 \mu_2}$ & $\sigma_{\mu_1 \mu_2 \mu_3}$ & $\sigma_{\mu_1 \mu_2 \mu_3 \mu_4}$ & $\sigma^{P-}_{\mu_1 \mu_2 \mu_3 \mu_4}$ \\
\midrule
Bootstrap & 0.518 & 5.291 & 4.180 & 6.987 & 4.638 & 6.113 & $\cdots$ \\
3j & 0.523 & 5.338 & 4.212 & 7.053 & 4.598 & 5.892 & 6.436 \\
Gaussian ($\xi = 0.9$) & 0.521 & 5.320 & 4.203 & 7.024 & 4.594 & 5.833 & 6.399 \\
Gaussian ($\xi = 1.0$) & 0.522 & 5.329 & 4.208 & 7.037 & 4.598 & 5.872 & 6.428 \\
Gaussian ($\xi = 1.1$) & 0.520 & 5.323 & 4.205 & 7.028 & 4.601 & 5.880 & 6.441 \\
Gaussian ($\xi = 1.2$) & 0.519 & 5.309 & 4.198 & 7.006 & 4.602 & 5.859 & 6.434 \\
Gaussian ($\xi = 1.3$) & 0.515 & 5.288 & 4.187 & 6.971 & 4.603 & 5.814 & 6.411 \\
\midrule
$\mathbb{Z}_2$-even & $\epsilon$ & $\epsilon'$ & $\epsilon''$ & $T_{\mu_1 \mu_2}$ & $T'_{\mu_1 \mu_2}$ & $\epsilon_{\mu_1 \mu_2 \mu_3 \mu_4}$ & $\epsilon'_{\mu_1 \mu_2 \mu_3 \mu_4}$ \\
\midrule
Bootstrap & 1.413 & 3.830 & 6.896 & 3.000 & 5.509 & 5.023 & 6.421 \\
3j & 1.418 & 3.860 & 6.955 & 3.000 & 5.565 & 4.954 & 6.319 \\
Gaussian ($\xi = 0.9$) & 1.415 & 3.849 & 6.930 & 3.000 & 5.546 & 4.905 & 6.306 \\
Gaussian ($\xi = 1.0$) & 1.416 & 3.854 & 6.943 & 3.000 & 5.554 & 4.939 & 6.318 \\
Gaussian ($\xi = 1.1$) & 1.414 & 3.850 & 6.934 & 3.000 & 5.548 & 4.949 & 6.325 \\
Gaussian ($\xi = 1.2$) & 1.411 & 3.841 & 6.916 & 3.000 & 5.533 & 4.936 & 6.327 \\
Gaussian ($\xi = 1.3$) & 1.406 & 3.827 & 6.887 & 3.000 & 5.510 & 4.908 & 6.326 \\
\bottomrule
\end{tabular*}
\end{table}

\clearpage

\section{Conclusions and discussion} \label{conclusion}

In this paper, we studied the LLL-based fuzzy-sphere regularization of the
3D Ising CFT from the viewpoint of matrix models. We introduced a class of
Hamiltonians specified by a positive function
$\mathcal{F}\big(k(k+1),s\big)$, which includes the original model of
Ref.~\cite{Zhu:2022gjc} and the simplified model of Ref.~\cite{He:2025ong}.
The models in this class possess
all the symmetries of
the original and simplified models.
We showed that every model in the
class can be reformulated as a vector
model, as a matrix--vector model in which bosonic Hermitian matrices couple
to fermionic vectors through commutators with the fuzzy-sphere background
$L_i$, and, after integrating out the fermionic vectors, as a matrix model.
The function $\mathcal{F}$ enters only through the quadratic term of the
bosonic matrices, whose propagator is proportional to $\mathcal{F}$.
For the Gaussian choice $\mathcal{F}_\xi$, the propagator is
proportional to the heat kernel
$e^{-\xi k(k+1)/2s}$. Exact diagonalization for 
$N\le16$ showed that the
low-energy spectra at criticality agree with the 3D Ising CFT for
$1.0\le\xi\le2.1$ for the original model and for $0.9\le\xi\le1.3$ for the
simplified model, up to finite-size
corrections. The loss of agreement outside these ranges, and in
particular the different behaviors 
of the class for the original model and that for the
simplified model, show that the agreement
is not automatic. 
These results provide matrix-model realizations of
the 3D Ising CFT.

We emphasize that the matrix--vector models and the matrix models are
written in a basis-independent form. The actions consist only of traces,
commutators with $L_i$, and functions of $(L^{\rm ad})^2$, and they are
covariant under the $U(N)$ transformation \eqref{U(N)_covariance}
with $\Phi \rightarrow U^{\dagger}\Phi U$, where $\Phi$ is $A_{\mu}$, 
$B_{\mu}$ or $C_{\mu}$. The
orbital basis and the $3j$ symbols, which characterize the LLL Hamiltonian,
do not appear, and the geometry of the sphere enters only through the
background $L_i$. This makes it natural to regard $L_i$ as a background,
for example a classical solution or the (local) minimum of quantum effective action
, of a more fundamental matrix model, and
suggests that other backgrounds may be considered in the same framework.

These matrix models differ from conventional field theories on the fuzzy
sphere such as \eqref{fuzzy MQM}, which generally suffer from UV/IR mixing.
In our models, the bosonic matrices have no time-derivative term and act as
auxiliary fields whose propagator suppresses the modes with
$k\gtrsim\sqrt{s/\xi}$, while the dynamical degrees of freedom are carried by
fermions in the fundamental representation. Since the models reproduce the
spectrum of the 3D Ising CFT, they do not suffer from UV/IR mixing, at least
for the low-lying spectrum at criticality examined here. It would be
important to examine other observables, such as correlation functions and
OPE coefficients, in the matrix-model language. We also note that matrix
models with vector degrees of freedom have appeared in the description of
quantum Hall states, such as the matrix Chern--Simons model
\cite{Susskind:2001fb,Polychronakos:2001mi}. 
In the matrix Chern--Simons model, the matrices are
dynamical and the vector represents boundary degrees of freedom, whereas in
our models the fermions themselves are the vectors and the matrices are
auxiliary fields.

An important question is which choices of $\mathcal{F}$ yield the 3D Ising
CFT at criticality. 
We see in the original and simplified
models and their Gaussian deformations
that the
relevant angular momenta are $k \lesssim O(\sqrt{s})$.
We expect that this generally holds,
since $k =O(\sqrt{s})$ implies that
the momentum
is comparable to the inverse of 
the magnetic length.
We thus conjecture that, in the large-$s$
limit, the relevant data is the scaling function $f(x)$ defined by
$\mathcal{F}\big(k(k+1),s\big)\to f(x)$ with $x=k^2/2s$, for which
$\mathcal{F}_{\rm orig}$ and $\mathcal{F}_\xi$ give $f(x)=e^{-x}$ and
$e^{-\xi x}$, respectively. 
Our results indicate that, at $N=16$, there is a
window of $\xi$ in which the Gaussian deformation gives the 3D Ising CFT, and
that the window depends on the form of the interaction. Since the effective
range of angular momenta $k\lesssim\sqrt{2s/\xi}$ shrinks for a fixed $N$ as
$\xi$ increases, the window in the limit $s\to\infty$ may be wider. More generally, an interesting class including the Gaussian choices is that in which $\mathcal{F}^{-1}$ is a polynomial or a series in
$(L^{\rm ad})^2$. For example, $\mathcal{F}=\big(1+\sum_{n}c_n(L^{\rm ad})^{2n}\big)^{-1}$
gives the quadratic term
$\frac12\Tr A_\mu^2-\frac{c_1}{2}\Tr[L_i,A_\mu]^2+\mbox{higher-order terms}$, which includes the terms
taking the form of the
mass and kinetic terms of conventional fuzzy-sphere field theories and
arises naturally when a massive field is integrated out. 
Whether this class of choices gives
the 3D Ising CFT can be examined by exact diagonalization in the same way as
in section \ref{deformation}.
Or we need to appeal to the Monte Carlo 
method discussed below, if larger $N$
is required.

It would also be interesting to determine whether models of this type
emerge as low-energy effective theories of more fundamental matrix models,
such as the BFSS, BMN, and type IIB matrix models, after integrating out ultraviolet degrees of
freedom around a background characterized by $L_i$. 
The fermionic vectors may also be obtained from matrices. 

In view of such an effective-theory interpretation, it is 
important to answer the above question
on the freedom of choices of $\mathcal{F}$. It is also suggestive
that the propagator of the
bosonic matrices takes the following form
\begin{equation}
  \big\langle\Phi_{km}(\tau)\Phi_{k'm'}(\tau')\big\rangle
  =(-1)^m\,\delta_{kk'}\delta_{m,-m'}\,
  \mathcal{F}\big(k(k+1),s\big)\,\delta(\tau-\tau')\ ,
\end{equation}
where $\Phi=\sum_{k,m}\Phi_{km}\hat{Y}_{km}$ for $\Phi=A_\mu,B_\mu$, or
$C_\mu$.
For the Gaussian choice, it is proportional to the heat kernel
$e^{-\xi k(k+1)/2s}$, which 
is
reminiscent of the exact renormalization group, with $2s/\xi$ playing the
role of an effective scale. Our observation that a change of $\xi$ can be
compensated by the tuning of $g_0$ and $g_1$, within the window found above,
is consistent with this interpretation. (For the studies of renormalizaion group in a scalar 
field theory on the fuzzy sphere, see \cite{Kawamoto:2015qla}.)

The basis-independent matrix-model formulation may also enable efficient numerical studies at larger $N$. Since the matrices need not be expanded in fuzzy spherical harmonics, Monte Carlo simulations of the matrix models, whose computational cost is expected to grow only polynomially in $N$, could reach system sizes beyond those accessible by exact diagonalization. In this aspect, it would be crucial to choose
$\mathcal{F}$ such that 
the computational cost is reduced. Such simulations would, however, have to deal with the sign problem, since the fermion determinant is not positive definite in general (see also Ref.~\cite{Hofmann:2023llr} for quantum Monte Carlo simulations of a fuzzy-sphere model).
There is a possibility that
the complex Langevin method
works well for the present matrix models with appropriate choices of
$\mathcal{F}$. (For complex Langevin simulations of matrix models, 
see \cite{Nishimura:2019qal,Anagnostopoulos:2020xai,Anagnostopoulos:2026qvz}.)

Finally, since the reformulation relies only on the Hamiltonian consisting of
LLL-projected density--density interactions, its extension to LLL
models for other 3D CFTs, such as free scaler CFT, the $O(3)$
Wilson--Fisher fixed point and the $SO(5)$ deconfined critical point, is
straightforward: the fermionic vectors acquire additional flavor indices, and the bosonic matrices are introduced for each
interaction channel. Other important directions include CFTs on other
manifolds such as $T^2, S^3, S^4$ and $\mathbb{CP}^2$, and more general quantum field theories.
We hope that such studies help clarify a mechanism by which ordinary quantum
field theories emerge from matrix models, distinct both from large-$N$
reduction, which is restricted to large-$N$ theories, and from
field theories on noncommutative spaces, which suffer from UV/IR mixing.

\section*{Acknowledgments}
We would like to thank the participants of the workshop 
``Discrete Approaches to the Dynamics of Fields and Space–Time 2026'' for discussions. Especially, we are 
grateful to 
Masafumi Fukuma, Kazuki Hasebe, Goro Ishiki, 
Hikaru Kawai, Kiyoharu Kawana, Jun Nishimura and Tamiaki Yoneya
for comments. 
A part of the numerical calculations in this paper was carried out on Heian at YITP in Kyoto University.
A. T.  was supported in part by JSPS KAKENHI Grant 
Number JP25K07319.
T. S. was supported by JST SPRING, Japan Grant Number JPMJSP2167.
ChatGPT (GPT-6) and Claude (Opus 5.5) were used to assist with code development and manuscript editing. 
The authors reviewed and validated all AI-assisted codes and wrote the manuscript themselves, incorporating AI suggestions at their own discretion.
The authors take full responsibility for the content of the paper.

\appendix

\section{Tuning the parameters $g_0$ and $g_1$} \label{g0_g1_tuning}

In this appendix, we describe the procedure used to tune $g_0$ and $g_1$ and
report the resulting values $g_0^c$ and $g_1^c$. The parameters are tuned
separately for each model and each system size $N=2s+1$, with $h=1$ and
$R^2=N$. Our procedure follows Ref.~\cite{Zhou:2025liv}. Since reaching
criticality imposes only one condition on $(g_0,g_1)$, the procedure selects
the point on the critical line at which the finite-size deviations of the
low-lying spectrum from the 3D Ising CFT are smallest.

For a given pair of trial parameters, we obtain the eigenvalues and
eigenstates by exact diagonalization, and identify the eigenstates
corresponding to the operators of the 3D Ising CFT with conformal dimensions
not exceeding 3,
\begin{equation}
  \sigma\ ,\ \partial_\mu\sigma\ ,\ \partial_\mu\partial_\nu\sigma\ ,\
  \Box\sigma\ ,\ \epsilon\ ,\ \partial_\mu\epsilon\ ,\ T_{\mu\nu}\ ,
  \label{operator_less_than_3}
\end{equation}
by the procedure described in Sec.~\ref{sec:numerical_procedure}. Let
$\Delta_O$ denote the numerical conformal dimension of the eigenstate
associated with the operator $O$, normalized so that $\Delta_T=3$. We define
\begin{align}
  &\bm{v}_{\rm num}=
  \left(\Delta_{\partial_\mu\sigma}-\Delta_\sigma\ ,\
  \Delta_{\partial_\mu\partial_\nu\sigma}-\Delta_\sigma\ ,\
  \Delta_{\Box\sigma}-\Delta_\sigma\ ,\
  \Delta_{\partial_\mu\epsilon}-\Delta_\epsilon\ ,\ 3\right)^{\rm T}\ ,\\
  &\bm{v}_{\rm ref}=(1,\ 2,\ 2,\ 1,\ 3)^{\rm T}\ .
\end{align}
At a critical point, $\bm{v}_{\rm num}$ should approach $\bm{v}_{\rm ref}$ as
$s\to\infty$. We therefore minimize the cost function
\begin{equation}
  f(g_0,g_1)
  =\min_a\big\lVert\bm{v}_{\rm ref}-a\,\bm{v}_{\rm num}\big\rVert^2
  =\lVert\bm{v}_{\rm ref}\rVert^2\sin^2\theta\ ,
  \label{cost}
\end{equation}
where $\theta$ is the angle between $\bm{v}_{\rm num}$ and $\bm{v}_{\rm ref}$.
The cost function vanishes when the two vectors are parallel. Without the
common fifth component, it would also vanish whenever the differences
between the dimensions of the primary operators and their descendants were
integer multiples of a common constant; the fifth component removes this
ambiguity. We minimize \eqref{cost} by the Nelder--Mead method, using the
implementation in the Julia package Optim.jl.

Table~\ref{table:optimization_nm16} lists the tuned values at $N=16$ for the
original and simplified
models. 
Note that for 
the Gaussian deformation of
the original model, $g_1^c$ decreases monotonically with $\xi$ and becomes
negative for $\xi\ge1.6$.
However, as mentioned below \eqref{U_k_modified}, the original physical meanings of $g_0$ and $g_1$ are lost because the potential $U(\Omega_{ab})$ in deformed Hamiltonians includes higher derivative terms. 
We thus consider that
the models with $\xi\ge1.6$
are also physically meaningful.

As a check of the procedure, we compare our result for the original model
with that of Ref.~\cite{Zhu:2022gjc}. The tuned values in
Table~\ref{table:optimization_nm16}, together with \eqref{Haldane_potential},
give $V_0\simeq1.458$ and $V_1\simeq0.3189$ (and $V_l=0$ for $l\ge2$) at
$N=16$. Rescaling $h$, $V_0$, and $V_1$ by a common factor of $3.16$ gives
$h=3.16$, $V_0\simeq4.61$, and $V_1\simeq1.01$, which agree within $3\%$ with
the finite-size-optimized values $h=3.16$, $V_0=4.75$, and $V_1=1$ of
Ref.~\cite{Zhu:2022gjc}, although the tuning procedures differ.

\begin{table}[t]
\centering
\caption{Tuned parameters $g_0^c$ and $g_1^c$ at $N=16$ with $h=1$. Left:
  the original model \eqref{H_original_c} (``3j'') and its Gaussian
  models \eqref{H_deformed_c}. Right: the simplified model
  \eqref{H_simple_3j_c} (``3j'') and its Gaussian models
  \eqref{H_simple_deformed_c}.  }
\label{table:optimization_nm16}
\begin{minipage}[t]{0.48\linewidth}
\centering
\begin{tabular}[t]{lll}
\toprule
Original model & \multicolumn{1}{c}{$g_0^c$} & \multicolumn{1}{c}{$g_1^c$} \\
\midrule
3j & $3.40160$ & $1.23267$ \\
Gaussian ($\xi = 0.8$) & $3.80472$ & $1.36620$ \\
Gaussian ($\xi = 0.9$) & $3.52260$ & $1.26336$ \\
Gaussian ($\xi = 1.0$) & $3.32926$ & $1.15555$ \\
Gaussian ($\xi = 1.1$) & $3.20517$ & $1.02750$ \\
Gaussian ($\xi = 1.2$) & $3.12263$ & $0.87209$ \\
Gaussian ($\xi = 1.3$) & $3.06701$ & $0.68613$ \\
Gaussian ($\xi = 1.4$) & $3.03695$ & $0.45797$ \\
Gaussian ($\xi = 1.5$) & $3.02233$ & $0.18865$ \\
Gaussian ($\xi = 1.6$) & $3.02598$ & $-0.13608$ \\
Gaussian ($\xi = 1.7$) & $3.03466$ & $-0.50313$ \\
Gaussian ($\xi = 1.8$) & $3.05020$ & $-0.92345$ \\
Gaussian ($\xi = 1.9$) & $3.07011$ & $-1.39069$ \\
Gaussian ($\xi = 2.0$) & $3.08402$ & $-1.88594$ \\
Gaussian ($\xi = 2.1$) & $3.08168$ & $-2.36594$ \\
Gaussian ($\xi = 2.2$) & $3.06155$ & $-2.81299$ \\
Gaussian ($\xi = 2.3$) & $3.02407$ & $-3.20679$ \\
Gaussian ($\xi = 2.4$) & $2.96499$ & $-3.51336$ \\
Gaussian ($\xi = 2.5$) & $2.89954$ & $-3.77275$ \\
\bottomrule
\end{tabular}
\end{minipage}\hfill%
\begin{minipage}[t]{0.48\linewidth}
\centering
\begin{tabular}[t]{lll}
\toprule
Simplified model & \multicolumn{1}{c}{$g_0^c$} & \multicolumn{1}{c}{$g_1^c$} \\
\midrule
3j & $4.64471$ & $1.22976$ \\
Gaussian ($\xi = 0.7$) & $2.86493$ & $0.63083$ \\
Gaussian ($\xi = 0.8$) & $3.52255$ & $0.83051$ \\
Gaussian ($\xi = 0.9$) & $4.20539$ & $1.05759$ \\
Gaussian ($\xi = 1.0$) & $4.91711$ & $1.30899$ \\
Gaussian ($\xi = 1.1$) & $5.66685$ & $1.58335$ \\
Gaussian ($\xi = 1.2$) & $6.42845$ & $1.87965$ \\
Gaussian ($\xi = 1.3$) & $7.20339$ & $2.19793$ \\
Gaussian ($\xi = 1.4$) & $8.01202$ & $2.53400$ \\
Gaussian ($\xi = 1.5$) & $8.81981$ & $2.89057$ \\
Gaussian ($\xi = 1.6$) & $9.68663$ & $3.26182$ \\
Gaussian ($\xi = 1.7$) & $10.56021$ & $3.65165$ \\
\bottomrule
\end{tabular}
\end{minipage}
\end{table}

\section{Details of the numerical results} \label{numerical results}

In this appendix, we provide further details of our numerical results. 

    \begin{figure}[tbp]
      \centering
      \includegraphics[width=0.48\textwidth]{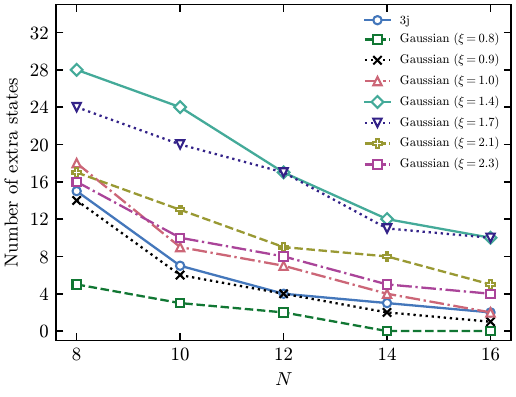}
      \hfill
      \includegraphics[width=0.48\textwidth]{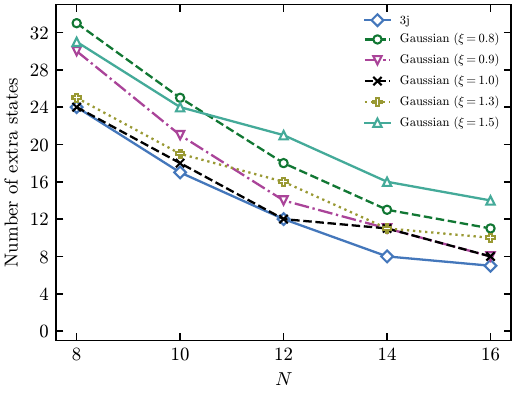}
      \caption{System-size dependence of the number of extra states with $\Delta < 7$ and $\ell\le4$ for each model. The left panel shows the results for the original model \eqref{H_original_c} and its
      Gaussian models \eqref{H_deformed_c}, while the right panel shows the results for
      the simplified model \eqref{H_simple_3j_c} and its Gaussian models \eqref{H_simple_deformed_c}.
      ``3j'' stands for the original model or the simplified model.}
      \label{fig:nm_dependence_extra_states}
    \end{figure}

\subsection{Extra states} \label{sec:n_dependence_extra_states}

In this Appendix, we examine extra states defined in section \ref{sec:numerical_procedure}. Their numbers as functions of $N$ are shown in Fig.~\ref{fig:nm_dependence_extra_states}. For every model, the number of extra states decreases with increasing $N$. At $N=16$, all extra states in all the models have conformal dimensions greater than or equal to 6.2. These observations are consistent with the models defined by \eqref{H_deformed_c}, \eqref{H_simple_3j_c}, and \eqref{H_simple_deformed_c} providing regularizations of the 3D Ising CFT.

\subsection{Detailed numerical results} \label{other_xi}

\begin{figure}[p]
      \centering
      \includegraphics[width=\textwidth]{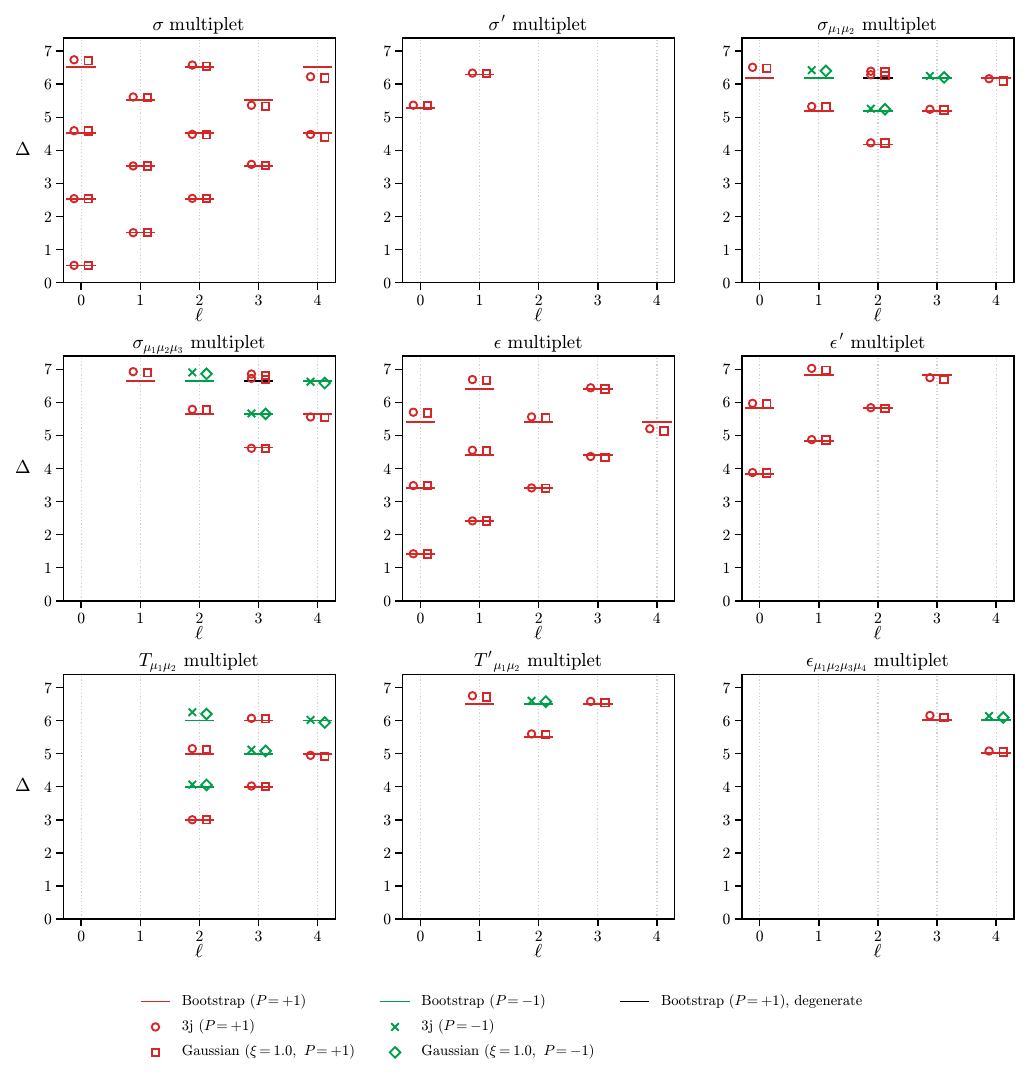}
      \caption{Conformal multiplets of primary operators that have descendants
  within $\Delta<7$ and $\ell\le4$ are shown for the original model and
  its Gaussian model at $\xi=1.0$.
   Each panel shows the spin $\ell$ on the
  horizontal axis and the conformal dimension $\Delta$ on the vertical axis.
  The conformal-bootstrap results of
  Refs.~\cite{Poland:2018epd,Simmons-Duffin:2016wlq} are shown
  by horizontal lines: 
  red lines denote parity-even non-degenerate operators,
  green lines denote parity-odd non-degenerate operators, and black lines
  denote parity-even twofold degenerate operators. For the original model (3j)
  \eqref{H_original_c}, the $N=16$ exact-diagonalization results are indicated
  by red circles (parity even) and green crosses (parity odd). For the
  Gaussian model \eqref{H_deformed_c} at $\xi=1.0$, they are indicated by red
  squares (parity even) and green diamonds (parity odd).}
      \label{fig:ising_multiplets_original_xi1.0}
    \end{figure}

\begin{figure}[p]
      \centering
      \includegraphics[width=\textwidth]{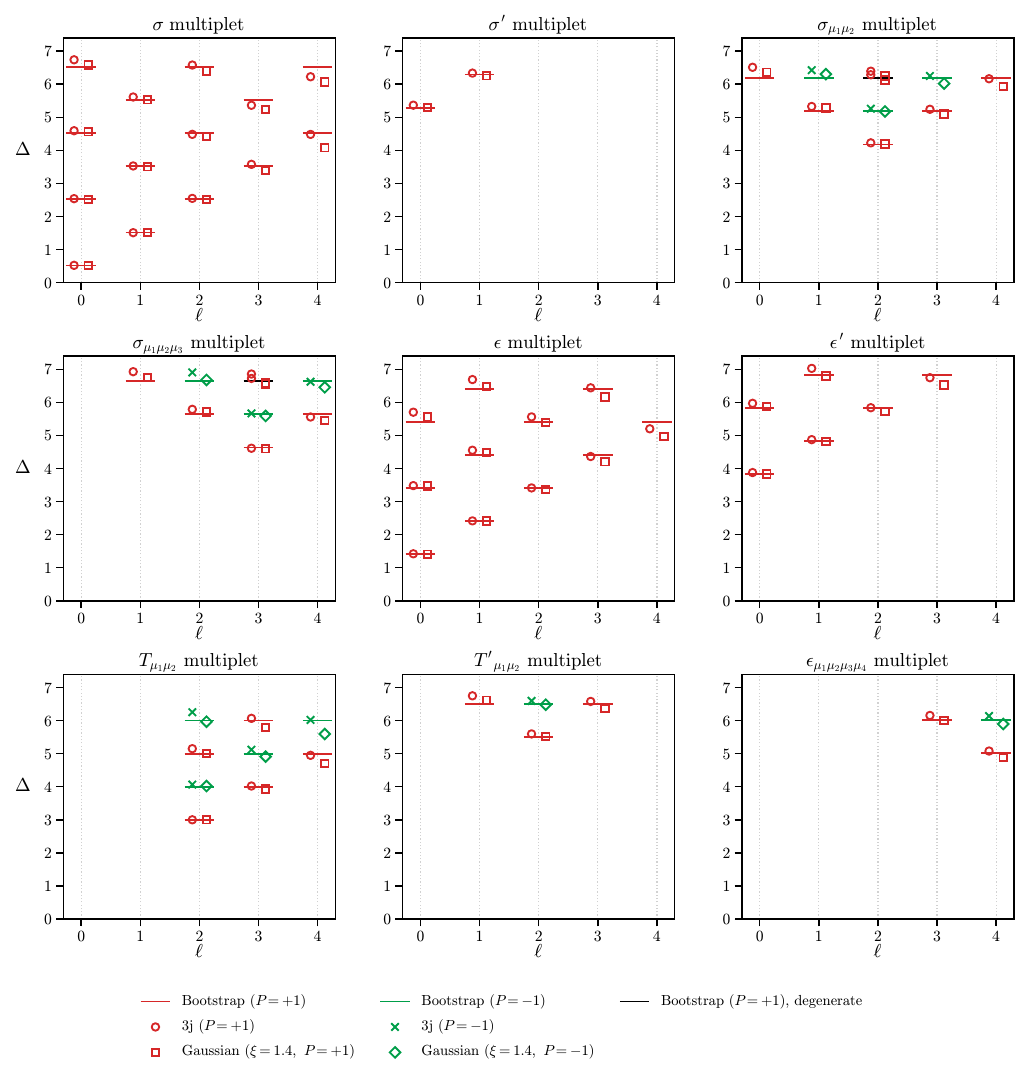}
      \caption{Conformal multiplets of primary operators that have descendants
  within $\Delta<7$ and $\ell\le4$ are shown for the original model and
  its Gaussian model at $\xi=1.4$.
   Each panel shows the spin $\ell$ on the
  horizontal axis and the conformal dimension $\Delta$ on the vertical axis.
  The conformal-bootstrap results of
  Refs.~\cite{Poland:2018epd,Simmons-Duffin:2016wlq} are shown
  by horizontal lines: 
  red lines denote parity-even non-degenerate operators,
  green lines denote parity-odd non-degenerate operators, and black lines
  denote parity-even twofold degenerate operators. For the original model (3j)
  \eqref{H_original_c}, the $N=16$ exact-diagonalization results are indicated
  by red circles (parity even) and green crosses (parity odd). For the
  Gaussian model \eqref{H_deformed_c} at $\xi=1.4$, they are indicated by red
  squares (parity even) and green diamonds (parity odd).}
      \label{fig:ising_multiplets_original_xi1.4}
    \end{figure}

\begin{figure}[p]
      \centering
      \includegraphics[width=\textwidth]{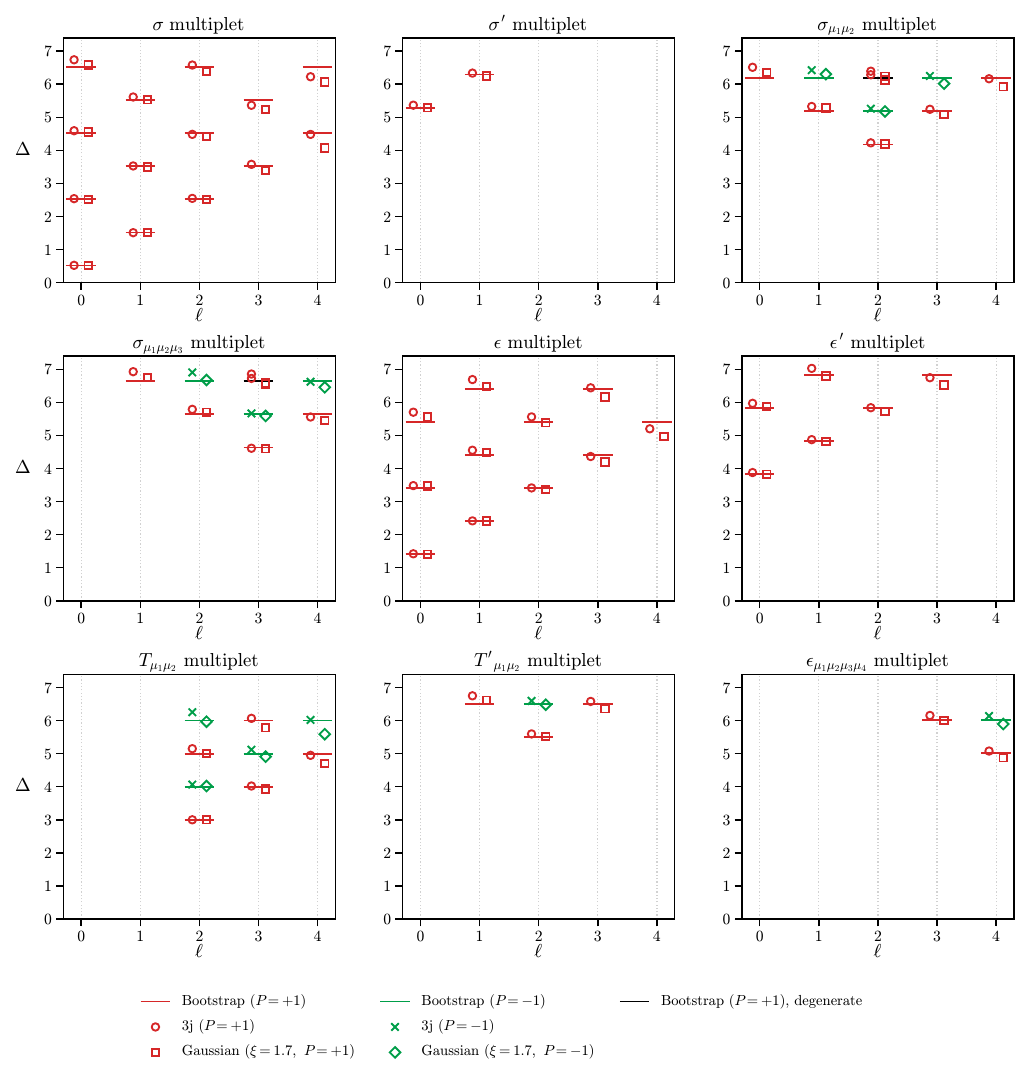}
      \caption{Conformal multiplets of primary operators that have descendants
  within $\Delta<7$ and $\ell\le4$ are shown for the original model and
  its Gaussian model at $\xi=1.7$.
   Each panel shows the spin $\ell$ on the
  horizontal axis and the conformal dimension $\Delta$ on the vertical axis.
  The conformal-bootstrap results of
  Refs.~\cite{Poland:2018epd,Simmons-Duffin:2016wlq} are shown
  by horizontal lines: 
  red lines denote parity-even non-degenerate operators,
  green lines denote parity-odd non-degenerate operators, and black lines
  denote parity-even twofold degenerate operators. For the original model (3j)
  \eqref{H_original_c}, the $N=16$ exact-diagonalization results are indicated
  by red circles (parity even) and green crosses (parity odd). For the
  Gaussian model \eqref{H_deformed_c} at $\xi=1.7$, they are indicated by red
  squares (parity even) and green diamonds (parity odd).}
      \label{fig:ising_multiplets_original_xi1.7}
    \end{figure}

\begin{figure}[p]
      \centering
      \includegraphics[width=\textwidth]{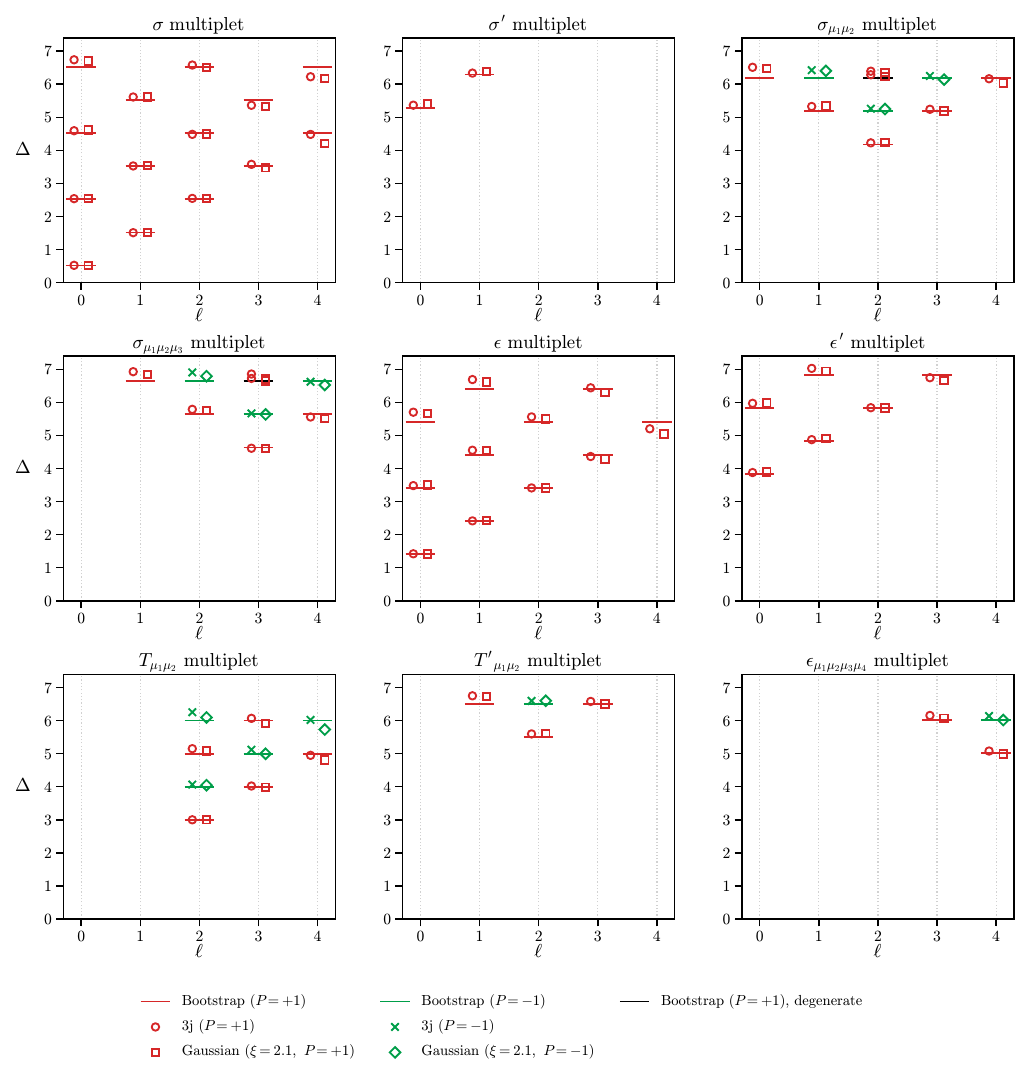}
      \caption{Conformal multiplets of primary operators that have descendants
  within $\Delta<7$ and $\ell\le4$ are shown for the original model and
  its Gaussian model at $\xi=2.1$.
   Each panel shows the spin $\ell$ on the
  horizontal axis and the conformal dimension $\Delta$ on the vertical axis.
  The conformal-bootstrap results of
  Refs.~\cite{Poland:2018epd,Simmons-Duffin:2016wlq} are shown
  by horizontal lines: 
  red lines denote parity-even non-degenerate operators,
  green lines denote parity-odd non-degenerate operators, and black lines
  denote parity-even twofold degenerate operators. For the original model (3j)
  \eqref{H_original_c}, the $N=16$ exact-diagonalization results are indicated
  by red circles (parity even) and green crosses (parity odd). For the
  Gaussian model \eqref{H_deformed_c} at $\xi=2.1$, they are indicated by red
  squares (parity even) and green diamonds (parity odd).}
      \label{fig:ising_multiplets_original_xi2.1}
    \end{figure}

\begin{figure}[p]
      \centering
      \includegraphics[width=\textwidth]{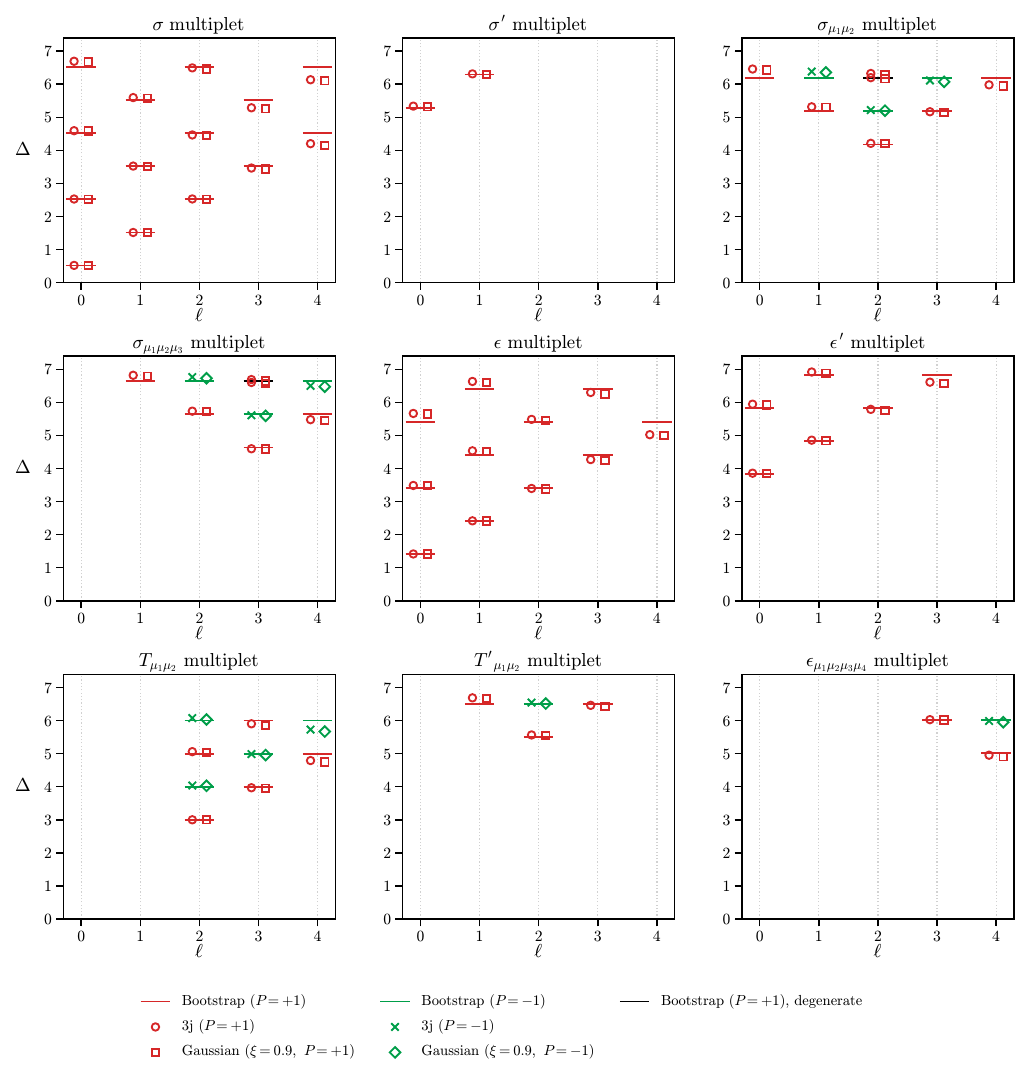}
      \caption{Conformal multiplets of primary operators that have descendants
  within $\Delta<7$ and $\ell\le4$ are shown for the simplified model and
  its Gaussian model at $\xi=0.9$.
   Each panel shows the spin $\ell$ on the
  horizontal axis and the conformal dimension $\Delta$ on the vertical axis.
  The conformal-bootstrap results of
  Refs.~\cite{Poland:2018epd,Simmons-Duffin:2016wlq} are shown
  by horizontal lines: 
  red lines denote parity-even non-degenerate operators,
  green lines denote parity-odd non-degenerate operators, and black lines
  denote parity-even twofold degenerate operators. For the simplified model (3j)
  \eqref{H_simple_3j_c}, the $N=16$ exact-diagonalization results are indicated
  by red circles (parity even) and green crosses (parity odd). For the
  Gaussian model \eqref{H_simple_deformed_c} at $\xi=0.9$, they are indicated by red
  squares (parity even) and green diamonds (parity odd).}
      \label{fig:ising_multiplets_simplified_xi0.9}
    \end{figure}

\begin{figure}[p]
      \centering
      \includegraphics[width=\textwidth]{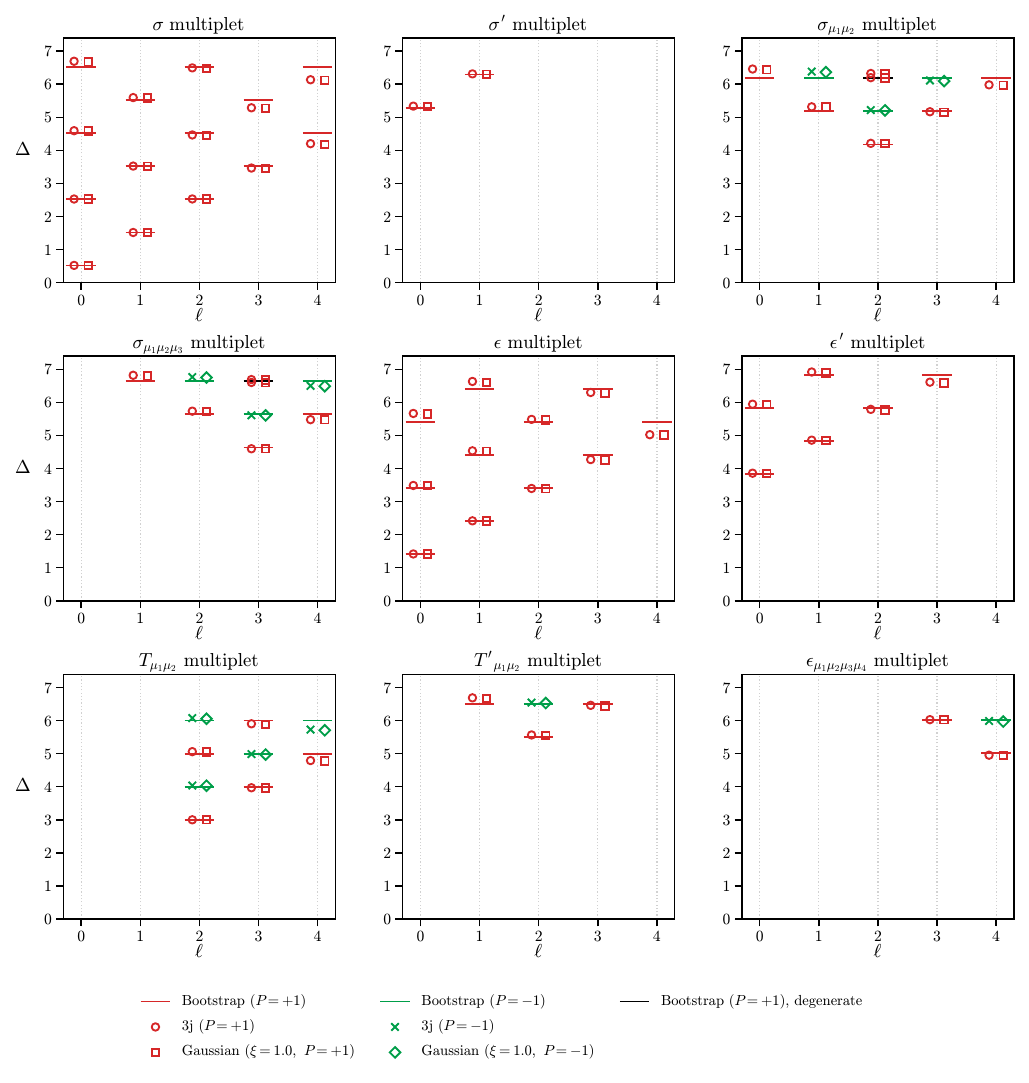}
      \caption{Conformal multiplets of primary operators that have descendants
  within $\Delta<7$ and $\ell\le4$ are shown for the simplified model and
  its Gaussian model at $\xi=1.0$.
   Each panel shows the spin $\ell$ on the
  horizontal axis and the conformal dimension $\Delta$ on the vertical axis.
  The conformal-bootstrap results of
  Refs.~\cite{Poland:2018epd,Simmons-Duffin:2016wlq} are shown
  by horizontal lines: 
  red lines denote parity-even non-degenerate operators,
  green lines denote parity-odd non-degenerate operators, and black lines
  denote parity-even twofold degenerate operators. For the simplified model (3j)
  \eqref{H_simple_3j_c}, the $N=16$ exact-diagonalization results are indicated
  by red circles (parity even) and green crosses (parity odd). For the
  Gaussian model \eqref{H_simple_deformed_c} at $\xi=1.0$, they are indicated by red
  squares (parity even) and green diamonds (parity odd).}
      \label{fig:ising_multiplets_simplified_xi1.0}
    \end{figure}

\begin{figure}[p]
      \centering
      \includegraphics[width=\textwidth]{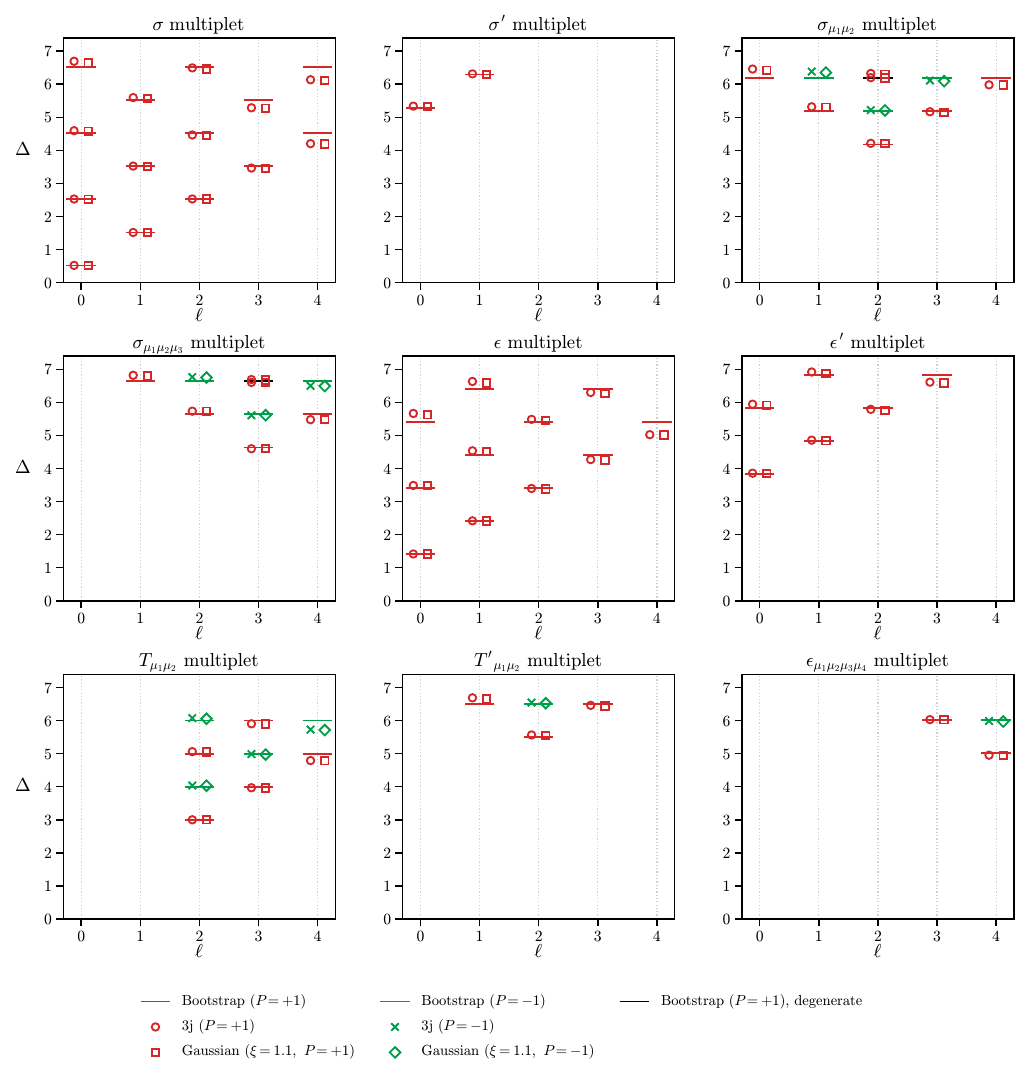}
      \caption{Conformal multiplets of primary operators that have descendants
  within $\Delta<7$ and $\ell\le4$ are shown for the simplified model and
  its Gaussian model at $\xi=1.1$.
   Each panel shows the spin $\ell$ on the
  horizontal axis and the conformal dimension $\Delta$ on the vertical axis.
  The conformal-bootstrap results of
  Refs.~\cite{Poland:2018epd,Simmons-Duffin:2016wlq} are shown
  by horizontal lines: 
  red lines denote parity-even non-degenerate operators,
  green lines denote parity-odd non-degenerate operators, and black lines
  denote parity-even twofold degenerate operators. For the simplified model (3j)
  \eqref{H_simple_3j_c}, the $N=16$ exact-diagonalization results are indicated
  by red circles (parity even) and green crosses (parity odd). For the
  Gaussian model \eqref{H_simple_deformed_c} at $\xi=1.1$, they are indicated by red
  squares (parity even) and green diamonds (parity odd).}
      \label{fig:ising_multiplets_simplified_xi1.1}
    \end{figure}

\begin{figure}[p]
      \centering
      \includegraphics[width=\textwidth]{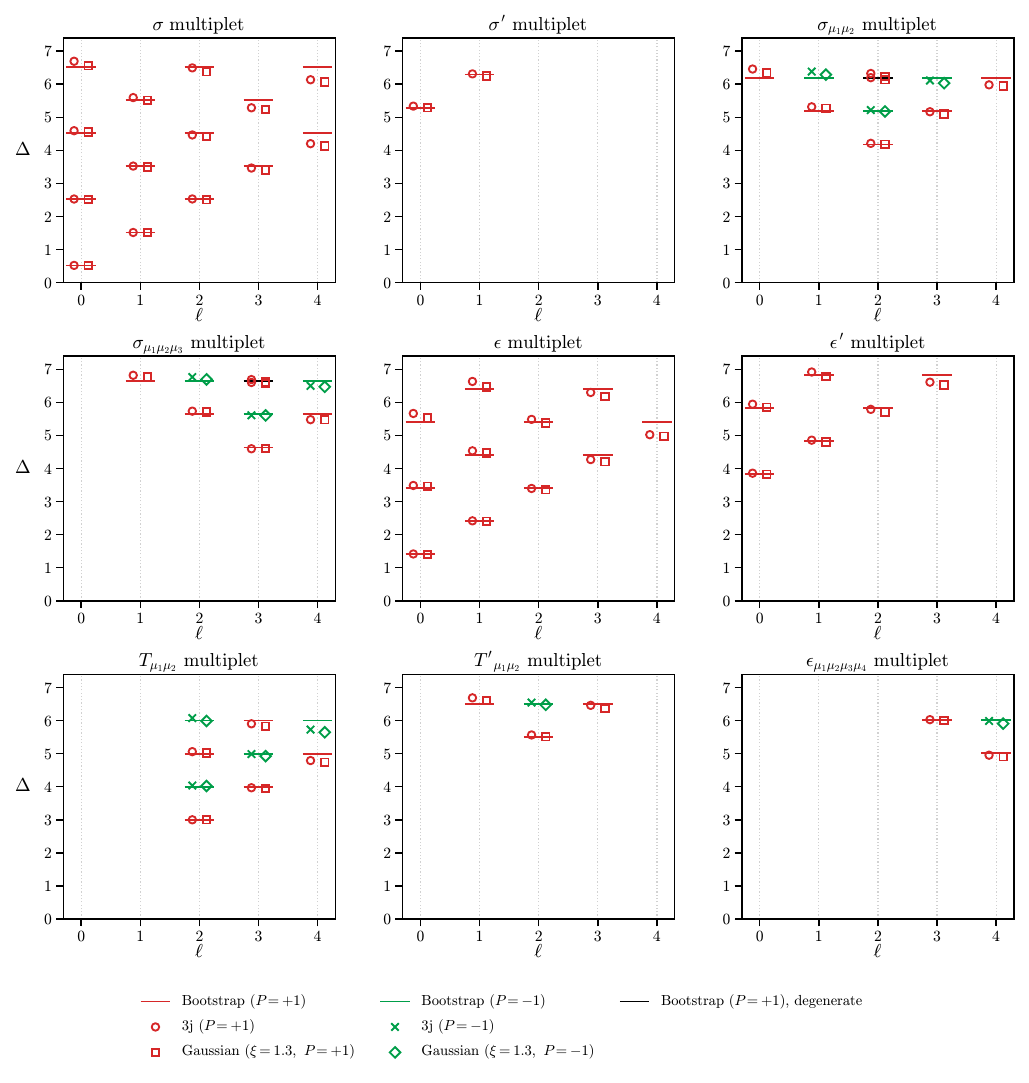}
      \caption{Conformal multiplets of primary operators that have descendants
  within $\Delta<7$ and $\ell\le4$ are shown for the simplified model and
  its Gaussian model at $\xi=1.3$.
   Each panel shows the spin $\ell$ on the
  horizontal axis and the conformal dimension $\Delta$ on the vertical axis.
  The conformal-bootstrap results of
  Refs.~\cite{Poland:2018epd,Simmons-Duffin:2016wlq} are shown
  by horizontal lines: 
  red lines denote parity-even non-degenerate operators,
  green lines denote parity-odd non-degenerate operators, and black lines
  denote parity-even twofold degenerate operators. For the simplified model (3j)
  \eqref{H_simple_3j_c}, the $N=16$ exact-diagonalization results are indicated
  by red circles (parity even) and green crosses (parity odd). For the
  Gaussian model \eqref{H_simple_deformed_c} at $\xi=1.3$, they are indicated by red
  squares (parity even) and green diamonds (parity odd).}
      \label{fig:ising_multiplets_simplified_xi1.3}
    \end{figure}

Here we present detailed $N=16$ exact-diagonalization results for the original model \eqref{H_deformed_c} and the simplified model \eqref{H_simple_deformed_c} at $\xi\ne1.2$.
For the original model, Figs.~\ref{fig:ising_multiplets_original_xi1.0}--\ref{fig:ising_multiplets_original_xi2.1} 
show the conformal multiplets at $\xi=1.0, 1.4, 1.7$ and $2.1$,
generalizing Fig.~\ref{fig:ising_multiplets_original_xi1.2}.
For the simplified model, Figs.~\ref{fig:ising_multiplets_simplified_xi0.9}--\ref{fig:ising_multiplets_simplified_xi1.3} 
show the conformal multiplets at $\xi=0.9, 1.0, 1.1$ and $1.3$,
generalizing Fig.~\ref{fig:ising_multiplets_simplified_xi1.2}.
The numerical multiplets in these figures agree well with the conformal-bootstrap data.

\clearpage

\clearpage

\clearpage

\bibliography{reference}
\bibliographystyle{utphys}
 
\end{document}